\documentclass[10pt,journal,compsoc,twoside]{IEEEtran}
\ifCLASSOPTIONcompsoc
  \usepackage[nocompress]{cite}
\else
  \usepackage{cite}
\fi
\ifCLASSINFOpdf
\else
\fi
\usepackage{amsmath,amssymb,mathrsfs,cases}
\usepackage{algorithm, algpseudocode}
\usepackage{cite}
\usepackage[level=1]{LatexInclusion/wgroup_message}  
\usepackage{color}  
\usepackage[usenames,dvipsnames]{xcolor}
\usepackage{colortbl,pgfplotstable}
\usepackage{psfrag}
\usepackage{graphicx}
\graphicspath{{Figures/}}

\newtheorem{remark}{Remark}

\pgfplotsset{compat=1.14}

\newcommand{\bd}{\begin{description}}
\newcommand{\ed}{\end{description}}
\newcommand{\be}{\begin{enumerate}}
\newcommand{\ee}{\end{enumerate}}
\newcommand{\bi}{\begin{itemize}}
\newcommand{\ei}{\end{itemize}}
\newcommand{\bl}{\begin{list}}
\newcommand{\el}{\end{list}}
\newcommand{\bt}{\begin{tabbing}}
\newcommand{\et}{\end{tabbing}}

\newcommand{\ist}{\hspace*{.3mm}}
\newcommand{\rmv}{\hspace*{-.3mm}}

\newcommand{\nn}{\nonumber}

\definecolor{BLUE}{rgb}{0,0,1}

\input{LatexInclusion/variables.tex}

\usepackage{acronym}
\acrodef{lids}[LIDS]{Laboratory for Information and Decision Systems}

\acrodef{nln}{network localization and navigation}
\acrodef{iid}[i.i.d.]{independent and identically distributed}
\acrodef{cdf}[CDF]{cumulative distribution function}
\acrodef{rf}[RF]{radio frequency}
\acrodef{ir}[IR]{infrared}
\acrodef{uwb}[UWB]{ultra-wideband}
\acrodef{lte}[LTE]{long-term evolution}
\acrodef{radar}[RADAR]{radio detection and ranging}
\acrodef{lidar}[LIDAR]{light detection and ranging}

\acrodef{wnl}[WNL]{wireless network localization}

\acrodef{cpnp}[CPNP]{conic programming-based node prioritization}
\acrodef{htna}[HTNA]{holistic threshold-based node activation}

\acrodef{fy}[FY]{fiscal year}

\acrodef{iot}[IoT] {Internet of Things}

\acrodef{ble}[BLE]{Bluetooth Low Energy}

\acrodef{pcb}[PCB]{printed circuit board}
\acrodef{ic}[IC]{integrated circuit}
\acrodef{phy}[PHY]{physical layer}
\acrodef{crc}[CRC]{cyclic redundancy check}
\acrodef{lr-wpans}[LR-WPANs]{low-rate wireless personal area networks}

\acrodef{gps}[GPS]{Global Positioning System}
\acrodef{gnss}[GNSS]{Global Navigation Satellite System}
\acrodefplural{gnss}[GNSS]{Global Navigation Satellite Systems}
\acrodef{imu}[IMU]{inertial measurement unit}
\acrodef{uav}[UAV]{unmanned aerial vehicles}

\acrodef{cvm}[CVM]{constant velocity model}

\acrodef{csi}[CSI]{channel state information}
\acrodef{csma}[CSMA]{carrier-sense multiple access}
\acrodef{ofdm}[OFDM]{orthogonal frequency division multiplexing}
\acrodef{snr}[SNR]{signal-to-noise-ratio}

\acrodef{socp}[SOCP]{second-order conic programming}
\acrodef{sdp}[SDP]{semidefinite programming}

\acrodef{ls}[LS]{least squares}
\acrodef{ml}[ML]{maximum likelihood}
\acrodef{lls}[LLS]{linear least squares}
\acrodef{nls}[LLS]{nonlinear least squares}
\acrodef{mmse}[MMSE]{minimum mean square error}
\acrodef{ekf}[EKF]{extended Kalman filter}
\acrodef{ukf}[UKF]{unscented Kalman filter}
\acrodef{bp}[BP]{belief propagation}
\acrodef{spbp}[SPBP]{sigma point belief propagation}
\acrodef{sp}[SP]{sigma point}
\acrodefplural{sp}[SPs]{sigma points}
\acrodef{nbp}[NBP]{nonparametric belief propagation}
\acrodef{gbp}[GBP]{Gaussian belief propagation}

\acrodef{pdf}[PDF]{probability density function}%
\acrodefplural{pdf}[PDFs]{probability density functions}

\acrodef{los}[LOS]{line-of-sight}
\acrodef{nlos}[NLOS]{non-line-of-sight}
\acrodef{toa}[TOA]{time-of-arrival}
\acrodef{tof}[TOF]{time-of-flight}
\acrodef{tdoa}[TDOA]{time-difference-of-arrival}
\acrodef{twr}[TWR]{two-way-ranging}
\acrodef{aoa}[AOA]{angle-of-arrival}
\acrodef{rss}[RSS]{received signal strength}
\acrodef{leo}[LEO]{localization error outage}
\acrodef{op}[OP]{outage probability}

\newacro{MLE}{maximum likelihood estimate}
\newacro{erc}[ERC]{equivalent ranging coefficient}

\acrodef{crlb}[CRLB]{Cram\'{e}r-Rao lower bound}
\acrodef{spe}[SPE]{squared position error}
\acrodef{speb}[SPEB]{squared position error bound}
\acrodef{cmpe}[CMPE]{covariance matrix of position estimate}
\acrodef{fim}[FIM]{Fisher information matrix}
\acrodef{ifim}[IFIM]{inverse Fisher information matrix}
\acrodef{efim}[EFIM]{equivalent Fisher information matrix}
\acrodef{iefim}[IEFIM]{inverse of the equivalent Fisher information matrix}
\acrodefplural{ifim}[IFIMs]{inverse Fisher information matrices}
\acrodef{pocs}[POCS]{projection onto convex sets}
\acrodef{mse}[MSE]{mean squared error}
\acrodef{rmse}[RMSE]{root-mean-square error}

\acrodef{nln}[NLN]{network localization and navigation}

\acrodef{cai}[CAI]{channel access indicator}
\acrodef{lmi}[LMI]{linear matrix inequality}

\usepackage{LatexInclusion/winsnotation}

\usepackage{standalone}
\usepackage{tikz} 
\usepackage{Figures/nodes}
\usepackage{Figures/factor}
\usepackage{Figures/architecture}

\usepackage[yyyymmdd,hhmmss]{datetime} 
\newdateformat{monthyeardate}{\monthname[\THEMONTH] \THEDAY, \THEYEAR} 

\newcolumntype{L}[1]{>{\raggedright\arraybackslash}p{#1}}
\newcolumntype{C}[1]{>{\centering\arraybackslash}p{#1}}
\newcolumntype{R}[1]{>{\raggedleft\arraybackslash}p{#1}}

\newcommand{\paperTitle}{Peregrine: Network Localization and Navigation\\ with Scalable Inference and Efficient Operation}

\begin{document}
%
\title{\paperTitle}
%
%
%
%

\author{
     Bryan~Teague,~\IEEEmembership{Member,~IEEE},
Zhenyu~Liu,~\IEEEmembership{Student~Member,~IEEE},
Florian~Meyer,~\IEEEmembership{Member,~IEEE}, 
Andrea~Conti,~\IEEEmembership{Senior Member,~IEEE},
and
Moe~Z.~Win,~\IEEEmembership{Fellow,~IEEE}
    \thanks{This research was supported, in part, by
  		the Office of Naval Research under Grant N00014-16-1-2141 and N62909-18-1-2017. 
	The material in this paper was presented, in part,
		at the IEEE Latin-American Conference on Communications, Guatemala City, Guatemala, November 2017.
    }    
    \thanks{
        B.\ Teague, Z.\ Liu, F.\ Meyer, and M.\ Z.\ Win are with
        the Laboratory for Information and Decision
        Systems (LIDS), Massachusetts Institute of Technology,
        Room 32-D666, 77 Massachusetts Avenue, Cambridge, MA 02139
        USA (e-mail: \{bteague, zliu14, fmeyer, moewin\}@mit.edu).
	}
    \thanks{
        A. Conti is with the Department of Engineering and CNIT, University of Ferrara, 44122 Ferrara, Italy (e-mail: a.conti@ieee.org).
    }
	
}

\IEEEtitleabstractindextext{%
\begin{abstract}

Location-aware networks will enable new services and applications in fields such as autonomous driving, smart cities, and the Internet-of-Things. One promising solution for ubiquitous localization is \ac{nln}, where devices form a network that cooperatively localizes itself, reducing the infrastructure needed for accurate localization. This paper introduces a real-time \ac{nln} system named Peregrine, which combines distributed \ac{nln} algorithms with commercially available \ac{uwb} sensing and communication technology. The Peregrine software application, for the first time, integrates three \ac{nln} algorithms to jointly perform the tasks of localization and network operation in a technology agnostic manner, leveraging both spatial and temporal cooperation. Peregrine hardware is composed of low-cost, compact devices that comprise a microprocessor and a commercial \ac{uwb} radio. This paper presents the design of the Peregrine system and characterizes the performance impact of each algorithmic component. Indoor experiments validate that our approach to realizing \ac{nln} is both reliable and scalable, and maintains sub-meter-level accuracy even in challenging indoor scenarios.

\acresetall

\end{abstract}

}
\acresetall		

\maketitle

\IEEEdisplaynontitleabstractindextext

%
\IEEEpeerreviewmaketitle


\section{Introduction}
\label{sec:introction}

\IEEEPARstart{P}{osition information} is becoming increasingly important \cite{WinConMazSheGifDarChi:J11}, enabling emerging applications in several areas including autonomy \cite{ThoWelLoiDanKum:17,WuChaYouBen:16,KarGus:J17}, crowdsensing \cite{YinRoyPoo:17,ZabCon:16,QuaChoCho:C14,ChoKimCho:C12,BarConWin:J17}, smart cities \cite{CarFosBelCorBorTalCur:13,ZanBuiCasVanZor:14,PasGiaBurChiFelLucZabFlaCasCriVerRobAndOre:18},  wireless sensor networks \cite{GezTiaGiaKobMolPooSah:05, PatAshKypHerMosCor:05, SimLeu:J14, ChiGioPao:18, CuoAbbCip:13, CuoCipAbb:09, PhoSoNiy:J18}, and the Internet-of-Things \cite{LuoHoaWanNiyKimHan:J16, CheThoJarLohAleLepBhuBupFerHonLinRuoKorKuu:17,WinMeyLiuDaiBarCon:J18,ZhaYanCheZhaGuoZha:16,NagZhaNec:17, SafKhaKarMou:18}. However, ubiquitous positioning remains extremely challenging in situations where \ac{gnss} signals are unavailable or unreliable. Such situations occur in many indoor environments, in cities and forests where the sky is occluded, and even in extra-terrestrial applications. In addition, \ac{gnss} signals do not provide the accuracy required by emerging applications such as motion tracking, autonomous navigation, and human-machine interfaces. The paradigm of \ac{nln} addresses this challenge by introducing cooperative measurements and information sharing in a \emph{localization network} \cite{WinSheDai:J18, SheWin:J10a,SheWymWin:J10, SheMazWin:J12,ConGueDarDecWin:J12}. \ac{nln} is enabled by \emph{node inference} and \emph{network operation} algorithms that are tailored to inexpensive hardware, and infrastructure or power limited applications \cite{SayTarKha:05,CafStu:98, LiuDaiWin:J18, ChoKum:03,PahLiMak:02}. 

\begin{figure}
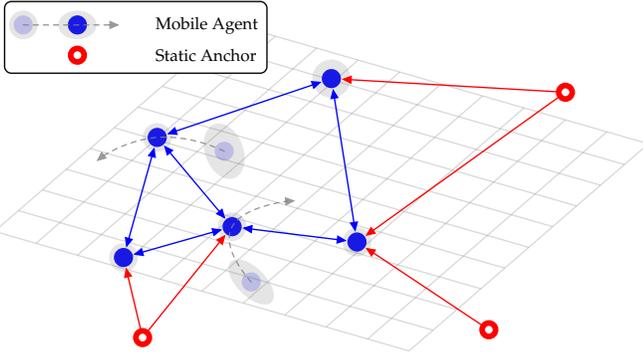

	\centering
	\includestandalone[width=\columnwidth]{nodes}
	\caption[Cooperative localization]{3-D \ac{nln} involving three anchors (red circles) and five agents (blue dots). The position uncertainties (gray ellipses) of all the agents and the time histories (gray lines) of two of the agents are also shown.}
	\label{fig:nodes}
\end{figure}

\ac{nln} infers the positions of mobile nodes using three sources of information (see Fig.~\ref{fig:nodes}) \cite{SheMazWin:J12}. First, agents perform pairwise measurements with anchors. Second, agents perform temporal filtering, also known as temporal cooperation using a motion model. Third, agents perform pairwise measurements with one another, which is referred to as spatial cooperation. The combination of temporal and spatial cooperation is referred to as spatiotemporal cooperation. In the example shown above, localization through traditional trilateration techniques would not be possible as no single agent is in range of three or more anchors. Only through spatiotemporal cooperation is localization possible.

This paper introduces Peregrine, a real-time system capable of demonstrating 3-D \ac{nln}. Each node in a Peregrine network contains fully distributed and asynchronous algorithms. Peregrine uses network operation algorithms to increase scalability and reduce infrastructure requirements. In particular, Peregrine uses both a \emph{node activation} algorithm to control channel access and a \emph{node prioritization} algorithm to select measurements. This combination of algorithms dynamically forms near-optimal network localization links. Localization is performed using inexpensive \ac{uwb} radio ranging modules and node inference algorithms, achieving sub-meter accuracies in challenging propagation environments. Note that \ac{nln} in 3-D is challenging compared to the 2-D case due to the fact that node inference suffers from the curse of dimensionality \cite{DauHua:03} and that certain network operation techniques designed specifically for 2-D scenarios are infeasible in 3-D \cite{DaiSheWin:J18}.

\subsection{Background and State of the Art}
\label{sub:background_and_state_of_the_art}

In the following, we provide background on the classes of algorithms and technologies relevant to this work. First, we discuss statistical inference strategies for cooperative localization. Second, we cover network operation algorithms designed specifically for localization networks. Third, we survey technologies and approaches for pairwise communication and measurements in ad hoc networks.

Algorithms for cooperative localization can be divided broadly into three categories: non-Bayesian, sequential Bayesian, and \ac{bp}-based techniques. Non-Bayesian algorithms fall into the categories of \ac{ls} and \ac{ml} \cite{Kay:B93}. \ac{ls} algorithms are suited for nonlinear measurement models and have low computational complexity. However, they may converge to local minima, do not naturally produce a confidence metric, typically require simultaneous measurements with three or four neighboring nodes, and do not naturally incorporate a statistical node motion model. \ac{ml} algorithms are guaranteed to find the most likely position, but they can be computationally complex as they search the entire {state} space.

Common sequential Bayesian estimation techniques include the \ac{ekf} \cite[Section 10.3]{ShaKirLi:B02}, the particle filter \cite[Section 4.3]{ThrFoxBur:B05}, and combinations thereof \cite{MazSheWin:J13}. These techniques are well understood and heavily utilized across a wide range of applications. Unfortunately, they are not suitable for the cooperative localization problem since they do not take the state uncertainty of neighboring nodes into account (if formulated in a distributed fashion) or they are not scalable (if formulated in a centralized fashion).

Bayesian algorithms relying on \ac{bp} represent the state-of-the-art in cooperative localization. Specifically, particle-based \ac{bp}, also known as \ac{nbp}, allows a distributed solution suited to nonlinear measurement models that takes into account the uncertainty of neighboring nodes\cite{IhlFisMosWil:05,WymLieWin:J09,LieFerSriWymWin:J12,MeyHliHla:J16}. Unfortunately, high-dimensional state estimation using \ac{nbp} is both computational intensive and results in very high communication overhead among nodes. In particular, due to the curse of dimensionality \cite{DauHua:03}, the number of particles needed to perform localization in 3-D can be infeasible for resource-limited devices. Recently, \ac{spbp} \cite{MeyHliHla:J14}, has been introduced, which uses a low-complexity approximation of \ac{bp} in nonlinear systems.

Another key aspect of \ac{nln} is algorithms for network operation \cite{WinDaiSheChrPoo:J18}. Key constraints on such systems are that nodes typically have limited battery life and that they need to share finite wireless resources for communication and measurements. This motivates the development of measurement selection \cite{BarGioWinCon:J15,DaiSheWin:J15,DaiSheWin:J15a} and channel access \cite{WanSheConWin:J17} algorithms to increase network lifetime, scalability, covertness, and localization performance. In particular, \textit{node activation} algorithms control channel access by determining \textit{when} a node should make measurements and exchange information. Node activation strategies have been developed to reduce delay \cite{GarMupWym:13}, communication overhead \cite{DwiZacAngHan:13}, and energy consumption \cite{GriBouPaz:10}. Recently, node activation algorithms have been developed to optimize channel access based on minimizing location error; i.e., those nodes that benefit the most from performing measurements should access the channel most frequently. \textit{Node prioritization} algorithms \cite{DaiSheWin:J18} determine \textit{which} nodes are best for pairwise ranging by selecting measurements based on the potential error reduction; i.e., the selected measurements should yield the maximum possible location information. 

\ac{nln} requires technologies for both pairwise measurements and communication between nodes in the network \cite{ConGueDarDecWin:J12,ConDarGueMucWin:J14}. Of the multitude of localization technologies that do not rely on \ac{gnss} \cite{Kap:B06,KarSto:B13,TitWes:J04,PatTorWanAli:J17}, those that use \iac{rf} link \cite{DarConFerGioWin:J09, WanUrrHanCab:11, WanCheCab:13} to simultaneously range and communication are extremely promising.
In particular, they are inexpensive, small, and do not require pointing, either by active involvement of a user or by the device itself. \ac{rf} links can pass through physical obstacles, smoke, or fog, which can occlude optical sensors. Unlike inertial sensors, \ac{rf} links can be used for absolute localization if anchors are present. Furthermore, data can be transferred between devices using the same link that is used to perform distance measurements. This facilitates the design of small, simple cooperative localization systems, where devices simultaneously range and share position information with one another. Such an architecture reduces reliance on infrastructure and improves overall network performance.

\ac{rf} technologies must combine wide signal bandwidth with low size, weight, power consumption, and cost in order to be suitable for widespread adoption for indoor localization. Large bandwidth allow multipath signal components to be resolved and mitigated, making them especially suitable for challenging propagation environments \cite{DarConFerGioWin:J09,WinSch:L98}. The two wireless technologies most often considered capable of producing accurate range measurements indoors are \ac{ofdm} \cite{VasKumKat:16, WanGaoMaoPan:J17, WanGaoMao:J17} and \ac{uwb} \cite{WinSch:L98,WinSch:L98a,WinSch:J00, CiaGup:09, RicCiaRos:16, GezTiaGiaKobMolPooSah:05,BarDaiConWin:J15, BarConDaiWin:L18}. {For \ac{ofdm} technologies, the carrier aggregation of multiple individual channels can be used to increase the total bandwidth for range measurements \cite{VasKumKat:16}. Such technique is typically processing intensive, and imperfections in the parameter estimation can degrade the ranging performance remarkably \cite{DaiLinWin:C17}.} In contrast, \ac{uwb} impulse radio provides {sufficiently high bandwidth for ranging} without the need for carrier aggregation. Moreover, chip-scale \ac{uwb} radios have recently become available, making it possible to develop high quality localization devices that are also inexpensive, small, and lightweight.

This paper combines these separate concepts and demonstrates their application to a real-time localization network. The specific contributions are covered below.

\subsection{Contribution and Organization of the Paper} 
\label{sub:contribution_and_paper_organization}

This paper presents a system for real-time 3-D \ac{nln}, which consists of the aforementioned \ac{nln} algorithms and inexpensive hardware nodes. In particular, we
\begin{itemize}
\item[$\bullet$] design low-complexity \ac{nln} algorithms for 3-D cooperative node inference, node activation, and node prioritization;
\vspace{1.5mm}

\item[$\bullet$] build a small, low-cost hardware node to demonstrate \ac{nln}; and
\vspace{1.5mm}

\item[$\bullet$] measure the individual contributions of node inference, node activation, and node prioritization to overall system performance.
\vspace{0.3mm}

\vspace{0mm}
\end{itemize}

Peregrine is the first system that implements cooperative node inference, node activation, and node prioritization for real-time 3-D localization in a hardware platform.\footnote{Peregrine system received an R\&D 100 Award \cite{RD100}; this paper presents distributed algorithms, efficient network operation, and network experimentation based on Peregrine devices.} 
Peregrine is a scalable system in the following ways. First, all algorithms for Peregrine are run in a distributed manner, and the complexity of algorithms for a specific agent increases only moderately with respect to the size of a subnetwork consisting of the agent and its neighbors. Second, systematically designed network operation algorithms are implemented in Peregrine so that contention for the wireless channel is carefully managed and communication resources are efficiently utilized even when the scale of the network increases.

The rest of the paper is organized as follows. Section~\ref{sub:system_model} introduces the system model. Sections~\ref{sec:inference} and \ref{sec:operation} present the node inference and network operation algorithms used in the Peregrine system, respectively. Section~\ref{sec:implementation} describes the implementation of the Peregrine system. Finally, Section~\ref{sec:results} presents our experimental results.

\mysubnote{Notation paragraph} \textit{Notation:} Random variables are displayed in sans serif, upright fonts; their realizations in serif, italic fonts. Vectors and matrices are denoted by bold lowercase and uppercase letters, respectively. For example, a random variable and its realization are denoted by $\rv{x}$ and $x$, respectively; a random vector and its realization are denoted by $\RV{x}$ and $\V{x}$, respectively; a matrix and a set are denoted by $\M{X}$ and $\Set{X}$, respectively. $f(\V x)$ denotes the \ac{pdf} $f_{\RV x}(\V x)$ of random vector $\RV x$, and $f(\V x | \V y)$ denotes the conditional \ac{pdf}  $f_{\RV x | \RV y}(\V x |\V y)$ of random vector $\RV x$ conditioned on random vector  $\RV y$; $\RV x \sim \mathcal N(\V \mu, \M \Sigma)$ denotes that random vector $\RV x$ follows the Gaussian distribution with mean $\V \mu$ and covariance matrix $\M \Sigma$.  The $m$-by-$m$ matrix of zeros is denoted by $\M{0}_{m \times m}$ and the $m$-by-$m$ identity matrix is denoted by $\M{I}_{m}$: the subscript is removed when the dimension of the matrix is clear from the context. ${\V{x}^{\text T}}$ and $\|{\V{x}}\|$ denote the transpose and the Euclidean norm of vector $\V x$, respectively; $\text{tr}\{\M X\}$ denote the trace of matrix $X$. The relation $\M X_1 \succeq \M X_2$ means that matrix $\M X_1 - \M X_2$ is positive semidefinite. Notation $\text{bdiag}\{\M X_1 ~ \M X_2 \ldots \M X_n\}$ denotes a block diagonal matrix with $\M X_1, \M X_2, \ldots, \M X_n$ on its main diagonal. 


\section{System Model}
\label{sub:system_model}

A decentralized network that consists of mobile agents with indices $\Nagt$ and static anchors with indices $\Nanc$ is considered.\footnote{In what follows, we will use the indices of nodes to denoted the corresponding nodes themselves.} In our model, time is discrete with time steps $n = 0,1,\dots$ and the duration of the intervals between time steps $n-1$ and $n$ is $t^{(n)}$. We denote the state of node $j \in \Nagt \cup \Nanc$ at time $n$ by vector $\RV{x}_{j}^{(n)} \in \mathbb{R}^{N_x}$, with ${N_x} \geq 3$ being its dimension. In particular, $\RV{x}_{j}^{(n)}$ can be written as $\RV{x}_{j}^{(n)} = \big[\, \RV{p}^{(n) \ist \text{T}}_{j} \ \RV v_j^{(n) \text T} \, \big]^{\text{T}}$, where $\RV{p}_{j}^{(n)} \in \mathbb{R}^{3}$ is the 3-D position of node $j$ at time step $n$, and $\RV v_j^{(n)}$ denotes other parameters related to position (e.g. velocity).

\subsection{Motion Model and Prior Distribution}\label{sec:mod_state}
For state evolution of mobile agent $j \in \Nagt$, we assume a linear motion model with Gaussian noise, i.e.,
\begin{equation}\label{eq:cvm}
\RV{x}_{j}^{(n)}  = \M{A}({t^{(n)}}) \ist \RV{x}_{j}^{(n-1)}  + \RV{w}_{j}^{(n)} \quad j \in \Nagt
\end{equation}
where $\M{A}({t^{(n)}})$ is the state evolution matrix; $\RV{w}_{j}^{(n)} \rmv\sim \Set{N}(\V{0}, \M{C}_w(t^{(n)}))$ is the \emph{driving noise} with generally arbitrary $\M{C}_w(t^{(n)})$, and it is assumed \ac{iid} across $j$ and $n$.

From the state-evolution model \eqref{eq:cvm}, we obtain the state-transition function $f\big(\V{x}_{j}^{(n)} \big | \V{x}_{j}^{(n-1)}\big)$ of agent $j \in \Nagt$. Similarly, for static anchors $j \in \Nanc$, we define $f\big(\V{x}_{j}^{(n)} \big | \V{x}_{j}^{(n-1)}\big) = \delta(\V{x}_{j}^{(n)} - \V{x}_{j}^{(n-1)})$, where $\delta(\cdot)$ is a Dirac delta function. At time step $n=0$, the states of all nodes $j \in \Nagt \cup \Nanc$ are assumed Gaussian distributed and statistically independent, i.e., $\RV{x}_{j}^{(0)} \sim \mathcal{N}\big(\V{\mu}_{j}^{(0)}, \M{C}_{j}^{(0)}\big)$. At time $n=0$ agents $j \in \Nagt$ are not localized which is expressed by the fact that the traces of their covariance matrices $\M{C}_{j}^{(0)}$ are large. Let us introduce the concatenated vector $\V{x}^{(0:n)} \triangleq [ \V{x}_{j}^{(n')} ]_{j \in \Nall, \, n'\in \{0,\ldots,n\}}$. The \textit{joint prior distribution} $f \big(\V{x}^{(0:n)}\big)$ can now be expressed as
\begin{equation}\label{eq:state_fact}
	f \big(\V{x}^{(0:n)}\big) = \prod_{j \in \Nagt \cup \Nanc} \rmv\rmv f\big(\V{x}_{j}^{(0)}\big) \prod_{n'=1}^{n} f\big(\V{x}_{j}^{(n')} \big | \V{x}_{j}^{(n'-1)}\big).
\end{equation}

\subsection{Measurement Model}\label{sec:mod_measure}

Let $\Nnb{j}^{(n)}$ be the set of nodes that are able to communicate and make distance measurements with agent $j$ at time step $n$. At time $n$, the control variable $m^{(n)}_{jk} \in \mathbb{N}_0$ determines the number of measurements performed by agent $j$ with each neighbor $k \in \Nnb{j}^{(n)}$. Similarly, the joint control vector of agent  $j$ is denoted as $\V{m}_{j}^{(n)} = [m_{jk}^{(n)}]_{k \in \Nnb{j}^{(n)}}$. We note that node $j$ and its neighbors in $\Nnb{j}^{(n)}$ compose a subnetwork, and we define the time steps to have exactly one agent performing measurements in each subnetwork at each time step.
In other words, we have $\|\V{m}^{(n)}_{j'}\| > 0$ for exactly one $j' \in \Nnb{j}^{(n)} \cup \{j\}$ and $\V{m}^{(n)}_{j'} = \V{0}$ otherwise. The measurement $\rv{z}_{jk}^{(n)}$ is the average of the $m^{(n)}_{jk}$ distance measurements that the agent $j$  performs with node $k$ for $\|{m}^{(n)}_{jk}\| > 0$. Specifically, $\rv{z}_{jk}^{(n)}$ is modeled as
\begin{align}\label{eq:meas_def}
	\rv{z}_{jk}^{(n)} = \|\RV{p}_{j}^{(n)} - \RV{p}_{k}^{(n)} \| + \rv{q}_{jk}^{(n)}.
\end{align}
The \textit{measurement noise} $\rv{q}_{jk}^{(n)}$ is distributed as $\rv{q}_{jk}^{(n)} \rmv\sim \Set{N}\big(0, \big[\sigma^{(n)}_{jk}\big]^2\big)$ and is assumed \ac{iid} across $j$, $k$, and $n$. The known measurement variance $\big[\sigma^{(n)}_{jk}\big]^2$ is given by $\big[\sigma^{(n)}_{jk}\big]^2 = \big(m^{(n)}_{jk} \xi^{(n)}_{jk}\big)^{-1} \rmv\rmv$, where $\xi^{(n)}_{jk}$ is the \ac{erc} \cite{DaiSheWin:J15a, DaiSheWin:J15,DaiSheWin:J18} that characterizes the channel quality between nodes $j$ and $k$ at time step $n$. 

For $\|\V{m}^{(n)}_j\| > 0$, the likelihood function $f(z_{jk}^{(n)}|\V{p}_{j}^{(n)},\V{p}_{k}^{(n)})$ can be directly obtained from the measurement model \eqref{eq:meas_def}. Let us introduce the set $\Set{M}_j^{(n)}$ that consists of all neighbors $k \in \Nnb{j}^{(n)}$ of agent $j \in \Nagt$ with $m^{(n)}_{jk} > 0$. In other words, $\Set{M}_j^{(n)}$ is the set of nodes with which agent $j$ makes measurements at time step $n$. Then, we define concatenated vectors $\V{z}^{(1:n)}$ as $\V{z}^{(1:n)}\triangleq \big[ z_{jk}^{(n')} \big]_{j \in \Nagt, \, k \in \Set{M}_j^{(n')}, \, n'\in \{1, 2, \ldots,n\}}$. We can now express the \textit{joint likelihood function} $f\left(\V{z}^{(1:n)} \big | \V{x}^{(1:n)} \right) $ as follows
\begin{align} \label{eq:likelihood_fact}
&f\big(\V{z}^{(1:n)} \big | \V{x}^{(1:n)}\big) \nn\\
&\hspace{12mm}= \prod_{n'=1}^{n}\prod_{j \in \Nagt} \prod_{k \in \Set{M}_j^{(n')}} f\big({z}_{jk}^{(n')} \big | \V{p}_{j}^{(n')}, \V{p}_{k}^{(n')}\big). \\
\nn\\[-12mm]
\nn
\end{align}
\vspace{-1mm}

\subsection{Problem Formulation} 
\label{sub:problem_formulation}

The following two tasks are performed locally at each agent $j \in \Nagt$:

\begin{enumerate}
\item 
Node Inference: The goal of node inference is to estimate the state of agent $j$ at each time step, incorporating distance measurements when they are made. We use a Bayesian estimation algorithm that exploits the statistical independence of driving noise in the state-evolution and measurement noise in the measurement model. As a result, we reduce the complexity and increase the scalability of node inference.
\vspace{1mm}

\item 
Network Operation: The goal of network operation is to reduce the localization error of the entire network and minimize the communication interference by controlling how each agent makes distance measurements. Network operation consists of two algorithms that are executed locally at the individual agents: node activation and node prioritization. In particular, node activation determines when agent $j$ should next access the channel, and node prioritization selects distance measurements agent $j$ should make with its neighbors when it does access the channel.

\end{enumerate}

Peregrine is designed to demonstrate \ac{nln} on small, low-power devices. Consequently, we assume limits on processing capability and communication resources (e.g., power, bandwidth, and channel access opportunity). The node inference and network operation algorithms described in the following sections are designed to require relatively little computation and communication overhead and to be executed in a distributed and scalable manner.

\section{Node Inference}\label{sec:inference}

In node inference, the 3-D location of mobile agents is determined from distance measurements with anchors and other mobile agents. Node inference is based on \ac{spbp} that enables spatiotemporal cooperation on resource-limited devices \cite{MeyHliHla:J14}. \ac{spbp} extends the \ac{sp} filter \cite{JulUhl:C97,WanMer:B01} to nonsequential Bayesian inference on loopy factor graphs. It is suitable for nonlinear problems, and avoids the communication and computation overhead of particle-based BP \cite{IhlFisMosWil:05}.

\subsection{MMSE Estimation and the \ac{bp} Algorithm}
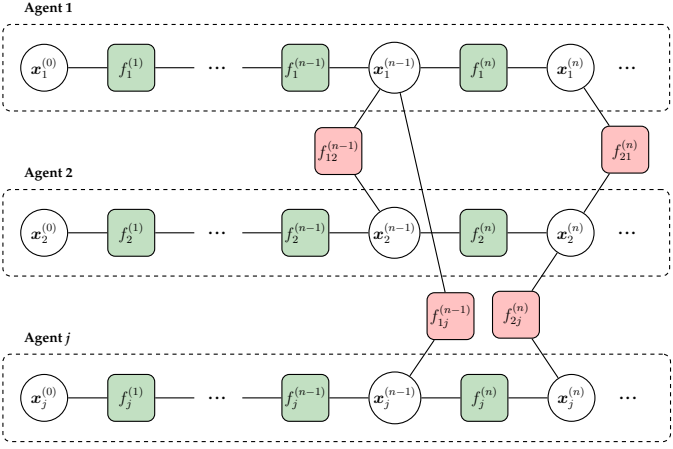
\begin{figure}
	\centering
	\begin{tikzpicture}[thick]  

  \definecolor{temporalcolor}{RGB}{194,224,194}
  \definecolor{spatialcolor}{RGB}{255,194,194}

  \node[latent]            (x_1^0)    {$\bm{x}_{1}^{(0)}$} ; %
  \factor[right=of x_1^0, fill=temporalcolor, opacity=1]       {f_1^1}     {$f_1^{(1)}$} {} {} ; %
  \node[const, right=of f_1^1]  (dots_11)    {\bf{\ldots}} ; %
  \factor[right=of dots_11, fill=temporalcolor, opacity=1]     {f_1^n-1}     {$f_1^{(n-1)}$} {} {} ; %
  \node[latent, right=of f_1^n-1] (x_1^n-1)  {$\bm{x}_{1}^{(n-1)}$} ; %
  \factor[right=of x_1^n-1, fill=temporalcolor, opacity=1]     {f_1^n}     {$f_1^{(n)}$} {} {} ; %
  \node[latent, right=of f_1^n] (x_1^n)    {$\bm{x}_{1}^{(n)}$} ; %
  \node[const, right=0.25cm of x_1^n]  (dots_12)    {\bf{\ldots}} ; %
  \node[const, above=0.5cm of x_1^0, minimum size=0cm]  (dummy_1)    {} ; %

  \factoredge {x_1^0}  {f_1^1}     {dots_11} ; %
  \factoredge {dots_11}  {f_1^n-1}     {x_1^n-1} ; %
  \factoredge {x_1^n-1}  {f_1^n}     {x_1^n} ; %

  \plate {plate1} { %
    (x_1^0)(f_1^1)(dots_11)(f_1^n-1)(x_1^n-1)(f_1^n)(x_1^n)(dots_12)   
  } {\bf{Agent 1}}; %

  \node[latent, below=3cm of x_1^0]                 (x_2^0)    {$\bm{x}_{2}^{(0)}$} ; %
  \factor[right=of x_2^0, fill=temporalcolor, opacity=1]       {f_2^1}     {$f_2^{(1)}$} {} {} ; %
  \node[const, right=of f_2^1]  (dots_21)    {\bf{\ldots}} ; %
  \factor[right=of dots_21, fill=temporalcolor, opacity=1]     {f_2^n-1}     {$f_2^{(n-1)}$} {} {} ; %
  \node[latent, right=of f_2^n-1] (x_2^n-1)  {$\bm{x}_{2}^{(n-1)}$} ; %
  \factor[right=of x_2^n-1, fill=temporalcolor, opacity=1]     {f_2^n}     {$f_2^{(n)}$} {} {} ; %
  \node[latent, right=of f_2^n] (x_2^n)    {$\bm{x}_{2}^{(n)}$} ; %
  \node[const, right=0.25cm of x_2^n]  (dots_22)    {\bf{\ldots}} ; %
  \node[const, above=0.5cm of x_2^0, minimum size=0cm]  (dummy_2)    {} ; %
  
  \factoredge {x_2^0}  {f_2^1}     {dots_21} ; %
  \factoredge {dots_21}  {f_2^n-1}     {x_2^n-1} ; %
  \factoredge {x_2^n-1}  {f_2^n}     {x_2^n} ; %

  \plate {plate2} { %
    (x_2^0)(f_2^1)(dots_21)(f_2^n-1)(x_2^n-1)(f_2^n)(x_2^n)(dots_22)   
  } {\bf{Agent 2}}; %

  \node[latent, below=3cm of x_2^0]                 (x_j^0)    {$\bm{x}_{j}^{(0)}$} ; %
  \factor[right=of x_j^0, fill=temporalcolor, opacity=1]       {f_j^1}     {$f_j^{(1)}$} {} {} ; %
  \node[const, right=of f_j^1]  (dots_j1)    {\bf{\ldots}} ; %
  \factor[right=of dots_j1, fill=temporalcolor, opacity=1]     {f_j^n-1}     {$f_j^{(n-1)}$} {} {} ; %
  \node[latent, right=of f_j^n-1] (x_j^n-1)  {$\bm{x}_{j}^{(n-1)}$} ; %
  \factor[right=of x_j^n-1, fill=temporalcolor, opacity=1]     {f_j^n}     {$f_j^{(n)}$} {} {} ; %
  \node[latent, right=of f_j^n] (x_j^n)    {$\bm{x}_{j}^{(n)}$} ; %
  \node[const, right=0.25cm of x_j^n]  (dots_j2)    {\bf{\ldots}} ; %
  \node[const, above=0.5cm of x_j^0, minimum size=0cm]  (dummy_j)    {} ; %

  \factoredge {x_j^0}  {f_j^1}     {dots_j1} ; %
  \factoredge {dots_j1}  {f_j^n-1}     {x_j^n-1} ; %
  \factoredge {x_j^n-1}  {f_j^n}     {x_j^n} ; %

  \factor[below left=1.05cm and 0.35cm of x_1^n-1, fill=spatialcolor, opacity=1] {f_12^n-1} {$f_{12}^{(n-1)}$}{}{}; %
  \factor[below right=1.05cm and 0.35cm of x_2^n-1, fill=spatialcolor, opacity=1] {f_1j^n-1} {$f_{1j}^{(n-1)}$}{}{}; %
  \factor[below right=1.05cm and 0.35cm of x_1^n, fill=spatialcolor, opacity=1] {f_21^n} {$f_{21}^{(n)}$}{}{}; %
  \factor[below left=1.05cm and 0.35cm of x_2^n, fill=spatialcolor, opacity=1] {f_2j^n} {$f_{2j}^{(n)}$}{}{}; %
  \factoredge {x_2^n-1} {f_12^n-1}   {x_1^n-1} ; %
  \factoredge {x_j^n-1} {f_1j^n-1}   {x_1^n-1} ; %
  \factoredge {x_1^n}   {f_21^n}     {x_2^n} ; %
  \factoredge {x_j^n}   {f_2j^n}     {x_2^n} ; %

  \plate {platej} { %
    (x_j^0)(f_j^1)(dots_j1)(f_j^n-1)(x_j^n-1)(f_j^n)(x_j^n)(dots_j2)   
  } {\bf{Agent \textit{j}}}; %

\end{tikzpicture}
	\caption{Part of the factor graph for \ac{nln}. A single agent accesses the channel at a specific time step $n$. We use the short notation $f^{(n)}_j \triangleq f(\V{x}_{j}^{(n)}|\V{x}_{j}^{(n-1)})$ and $f_{jk}^{(n)} \triangleq f(z_{jk}^{(n)}|\V{x}_{j}^{(n)},\V{x}_{k}^{(n)})$.}
	\label{fig:factor_graph}
\end{figure}

The \ac{mmse} estimator is adopted to determine the state of an agent. In particular, the \ac{mmse} estimate $\hat{\V{x}}^{(n)}_{j}$ of $\RV{x}^{(n)}_{j}$ based on $\V{z}^{(1:n)}$ is given by \cite{Kay:B93}

\begin{equation}
\hat{\V{x}}^{(n)}_{j} \,\triangleq \int \V{x}^{(n)}_j f(\V{x}^{(n)}_j|\V{z}^{(1:n)}) \ist d\V{x}^{(n)}_{j} \,.
\label{eq:mmse_slat}
\end{equation}
The MMSE estimate is based on the marginal posterior \ac{pdf} $f(\V{x}_{j}^{(n)}|\V{z}^{(1:n)})$, which is obtained by integrating the joint posterior \ac{pdf} $f(\V{x}^{(0:n)}|\V{z}^{(1:n)})$. However, direct marginalization of $f(\V{x}^{(0:n)}|\V{z}^{(1:n)})$ is infeasible because it relies on nonlocal information and involves integration in spaces whose dimension grows with time and network size. An efficient method to compute the marginal posterior \ac{pdf} is the \ac{bp} algorithm. The factor graph is obtained according to the joint posterior \ac{pdf} $f(\V{x}^{(0:n)}|\V{z}^{(1:n)})$. In particular, using Bayes' rule as well as \eqref{eq:state_fact} and \eqref{eq:likelihood_fact}, $f(\V{x}^{(0:n)}|\V{z}^{(1:n)})$ can be factored as
\begin{align}\label{eq:joint_posterior}
	f(\V{x}^{(0:n)}|\V{z}^{(1:n)}) &\propto f(\V{x}^{(0:n)}) f(\V{z}^{(1:n)}|\V{x}^{(1:n)}) \nn\\ 
	&=\prod_{j \in \Nagt \cup \Nanc} f(\V{x}_{j}^{(0)}) 	\prod_{n' = 1}^n f(\V{x}_{j}^{(n')}|\V{x}_{j}^{(n'-1)}) \nn\\
	&\hspace{7.3mm} \times \rmv \prod_{k \in \Set{M}_j^{(n')}} \rmv f(z_{jk}^{(n')}|\V{p}_{j}^{(n')},\V{p}_{k}^{(n')}) \\[-7mm]
	\nn
\end{align}
where we formally clarify measurement set $\Set{M}_j^{(n)} = \varnothing$ for all anchors $j \in \Nanc$. The corresponding factor graph is shown in Fig.~\ref{fig:factor_graph}. In contrast to the standard factor graph for cooperative localization \cite{WymLieWin:J09}, our factor graph in Fig.~\ref{fig:factor_graph} does not have any loops at a single time step $n$. This is related to the fact that only one agent in each subnetwork $\{j\} \cup \Nnb{j}$ is able to access the channel and thus a single agent is involved in all range measurements performed at time $n$.

The result of the \ac{bp} algorithm depends on the order in which the messages are computed. When the factor graph is loopy, there is no fixed order in which messages should be computed, and different orders may result in different approximates of the marginal posterior \acp{pdf}.
When used for real-time navigation, the order is defined by passing messages only forward in time \cite{WymLieWin:J09}.
Exchanging \ac{bp} messages \cite{KscFreLoe:01} using this order on the factor graph in Fig.~\ref{fig:factor_graph} results in the following approximate marginal posterior \acp{pdf} $b(\V{x}_{j}^{(n)})$, also known as ``beliefs'', for agent $j \in \Nagt$ at time $n$,
\begin{align}
	b(\V{x}_{j}^{(n)}) \propto \phi_{\rightarrow n}(\V{x}_{j}^{(n)}) \prod_{k \in \Set{M}^{(n)}_j} \phi_{k \rightarrow j}(\V{x}_{j}^{(n)}) \label{eq:bp_marg} 
\end{align}
with the ``prediction message''
\begin{align}
	\phi_{\rightarrow n}(\V{x}_{j}^{(n)}) = \int f(\V{x}_{j}^{(n)}|\V{x}_{j}^{(n-1)})\,
	b(\V{x}_{j}^{(n-1)})\, \mathrm{d}\V{x}_{j}^{(n-1)}\label{eq:bp_pred}
\end{align}
and the ``measurement message''
\begin{align}
	\phi_{k \rightarrow j}(\V{x}_{j}^{(n)}) &= \int f(z_{jk}^{(n)}|\V{p}_{j}^{(n)},\V{p}_{k}^{(n)}) \phi_{\rightarrow n}(\V{x}_{k}^{(n)})\, \mathrm{d}\V{x}_{k}^{(n)}\nn\\
	&= \int f(z_{jk}^{(n)}|\V{p}_{j}^{(n)},\V{p}_{k}^{(n)}) \phi^{\hspace{.15mm} \text{p}}_{\rightarrow n}(\V{p}_{k}^{(n)})\, \mathrm{d}\V{p}_{k}^{(n)}\rmv\rmv.\label{eq:bp_meas}
\end{align}
The ``prediction message'' $\phi_{\rightarrow n}(\V{x}_{j}^{(n)})$ is calculated from the state transition \ac{pdf} and the previous belief. The ``measurement message'' $\phi_{k \rightarrow j}(\V{x}_{j}^{(n)})$ is based on a pairwise distance measurement between agent $j \in \Nagt$ and node $k \in \Set{M}_j^{(n)}$ at time $n$. According to \eqref{eq:bp_meas}, $\phi_{k \rightarrow j}(\V{x}_{j}^{(n)})$ involves position information of the neighboring node $k \in \Set{M}_j^{(n)}$ in the form of the ``prediction message''  $\phi^{\text{p}}_{\rightarrow n}(\V{p}_{k}^{(n)}) = \int \phi_{\rightarrow n}(\V{x}_{k}^{(n)}) \ist \mathrm{d} \V{v}_{k}^{(n)}$ related to its position. Therefore, these prediction messages need to be exchanged among neighboring nodes during the process of making distance measurements for belief updates. The exact beliefs are difficult to compute, due to the integration of \eqref{eq:bp_pred} and \eqref{eq:bp_meas} as well as the message multiplication in \eqref{eq:bp_marg}. For this reason, we use the \ac{spbp} algorithm to compute an approximate representation of these beliefs.

\begin{remark}
An alternative node inference approach to this BP algorithm is to infer the marginal posterior $f \big(\V{x}^{(n)}|\V{z}^{(1:n)}\big)$ by means of sequential Bayesian estimation and calculate an MMSE estimate $\hat{\V{x}}^{(n)}$ of the joint agent state $\RV{x}^{(n)}$ i.e., the concatenation of $\RV{x}_j^{(n)}$ over all $j$. However, as analyzed in \cite{MeyHliHla:J16}, this approach is not scalable in the network size since the dimension of the state to be estimated increases with the number of agents in the network. Furthermore, this approach is not amendable for distributed implementation. In contrast, the inference approach presented in this section is scalable to hundred of devices as demonstrated by simulation in \cite{MeyHliHla:J16}.
\end{remark}

\subsection{Sigma Point Belief Propagation}\label{sec:spbp}
In the \ac{spbp} algorithm, the belief $b(\V{x}_{j}^{(n)})$ of  any agent $j \in \Nagt$ is represented at each time $n$ by a mean $\V{\mu}_{j}^{(n)}$ and a covariance matrix $\M{C}_{j}^{(n)}$, which are updated at each time step.

At each time step, the \ac{spbp} algorithm potentially consists of two phases. In the first phase, the prediction message $\phi_{\rightarrow n}(\V{x}_{j}^{(n)})$ is evaluated using the state transition model \eqref{eq:cvm}. Since \eqref{eq:cvm} is a linear model of the state with Gaussian noise, from the mean $\V{\mu}_{j}^{(n)}$ and the covariance matrix $\M{C}_{j}^{(n)}$  representing $b(\V{x}_{j}^{(n-1)})$, we can directly get the predicted  mean $\tilde{\V{\mu}}_{j}^{(n)}$ and the predicted covariance matrix $\tilde{\M{C}}_{j}^{(n)}$ representing $\phi_{\rightarrow n}(\V{x}_{j}^{(n)})$, according to
\begin{align}
\tilde{\V{\mu}}_{j}^{(n)} &= \M{A}(t^{(n)}) \ist \V{\mu}_{j}^{(n-1)}  \nn\\
\tilde{\M{C}}_{j}^{(n)} &= \M{A}(t^{(n)}) \ist \M{C}_{j}^{(n-1)} \M{A}^{\mathrm{T}}(t^{(n)}) +   \ist \M{C}_w(t^{(n)}). \label{eq:meanCovariance}
\qquad
\end{align}
In the second phase of the \ac{spbp} algorithm, the distance measurements are incorporated to obtain a mean $\V{\mu}_{j}^{(n)}$ and a covariance matrix $\M{C}_{j}^{(n)}$ representing the updated belief $b(\V{x}_{j}^{(n)})$ by reformulating the \ac{bp} algorithm in a higher-dimensional state space \cite{MeyHliHla:J14}. Such reformulation enables solving the message passing equations \eqref{eq:bp_marg}--\eqref{eq:bp_meas} in an efficient manner even with nonlinear measurement model \eqref{eq:meas_def}. In particular, assuming $\Set{M}_j^{(n)} \!=\! \{k_1, k_2, \dots, k_{|\Set{M}_j^{(n)}|} \}$, we obtain a stacked (dimension-augmented)  vector $\overline{\V{x}}^{(n)}_{j}$ as 
\begin{align*}
\overline{\V{x}}^{(n)}_{j} &\triangleq \big[ \V{x}_{j}^{(n)\text{T}} \V{p}^{(n)\text{T}}_{\sim j} \big]^{\text T}
\end{align*}
where $\V{p}^{(n)}_{\sim j}$ is
\begin{align*}
\V{p}^{(n)}_{\sim j} &\triangleq \big[ \V{p}_{k_{1}}^{(n)\text{T}} \;\,  \V{p}_{k_{2}}^{(n)\text{T}} \cdots\ist \V{p}_{k_{|\Set{M}^{(n)}_j|}}^{(n)\text{T}} \big]^{\text{T}}.
\end{align*}%
We also obtain a stacked measurement vector $\overline{\V{z}}^{(n)}_{j}$ as
\begin{align*}
\overline{\V{z}}^{(n)}_{j} &\triangleq \big[ z_{jk_{1}}^{(n) } \;\, z_{jk_{2}}^{(n) } \cdots\ist z_{jk_{|\Set{M}^{(n)}_j|}}^{(n) } \big]^{\text{T}}.
\end{align*}
Message passing equations \eqref{eq:bp_marg}--\eqref{eq:bp_meas} can now be reformulated as follows:
\begin{align}\label{eq:bp_hd}
	b(\V{x}_{j}^{(n)}) &= \int b(\overline{\V{x}}_{j}^{(n)})\, \mathrm{d}\V{p}_{\sim j}^{(n)}
\end{align}
where $b(\overline{\V{x}}_{j}^{(n)})$ is defined as
\begin{align}\label{eq:bp_hd_joint}
	b(\overline{\V{x}}_{j}^{(n)}) &\propto   f(\overline{\V{x}}_{j}^{(n)}) \, f(\overline{\V{z}}_{j}^{(n)}|\overline{\V{x}}_{j}^{(n)})
\end{align}	
with
\begin{align}
	f(\overline{\V{x}}_{j}^{(n)}) &= \phi_{\rightarrow n}(\V{x}_{j}^{(n)}) \prod_{k \in \Set{M}^{(n)}_j} \phi_{\rightarrow n}(\V{x}_{k}^{(n)}) \nn\\
	f(\overline{\V{z}}_{j}^{(n)}|\overline{\V{x}}_{j}^{(n)}) &= \prod_{k \in \Set{M}^{(n)}_j} f(z_{jk}^{(n)}|\V{p}_{j}^{(n)},\V{p}_{k}^{(n)}).\label{eq:bp_hd_meas}
\end{align}
Equations \eqref{eq:bp_hd}, \eqref{eq:bp_hd_joint}, and \eqref{eq:bp_hd_meas} can be interpreted as follows. First, $b(\overline{\V{x}}_{j}^{(n)})$ can be seen as the result of a Bayesian update step on the augmented vector, with the prior $f(\overline{\V{x}}_{j}^{(n)})$ and the likelihood function $f(\overline{\V{z}}_{j}^{(n)}|\overline{\V{x}}_{j}^{(n)})$. Second, computation of $b({\V{x}}_{j}^{(n)})$ from $b(\overline{\V{x}}_{j}^{(n)})$ can be interpreted as a marginalization step. 

With the above reformulation, the updated mean $\V{\mu}_{j}^{(n)}$ and covariance $\M{C}_{j}^{(n)}$ corresponding to the belief $b(\V{x}_{j}^{(n)})$ of agent $j \in \Nagt$ can be calculated as in the following \cite{MeyHliHla:J14}.
\begin{enumerate}
\item  First, we obtain a mean vector $\overline{\V \lambda}_j^{(n)}$ and a covariance matrix $\overline{\M {\Sigma}}_j^{(n)}$ corresponding to the ``prior'' $f(\overline{\V{x}}_{j}^{(n)})$ 
as
\begin{align}
\overline{\V \lambda}_j^{(n)} &\triangleq \big[ \tilde{\V{\mu}}_{j}^{(n) \ist \text{T}} \;\, \V{\mu}_{\text{p},k_1}^{(n) \ist \text{T}} \;\,  \V{\mu}_{\text{p},k_2}^{(n) \ist \text{T}} \cdots\ist \V{\mu}_{\text{p}, k_{|\Set{M}^{(n)}_j|}}^{(n) \ist \text{T}} \big]^{\text{T}} \rmv, \nn\\[1mm]
\overline{\M {\Sigma}}_j^{(n)} &\triangleq \mathrm{bdiag} \Big\{ \tilde{\M{C}}_{j}^{(n)} \ist\ist \M{C}_{\text{p},k_1}^{(n)} \ist\ist \M{C}_{\text{p},k_2}^{(n)} \cdots\ist\ist \M{C}_{\text{p},k_{|\Set{M}^{(n)}_j|}}^{(n)} \Big\} \ist \label{eq:stackedMeanCov}
\end{align}
where $\V{\mu}_{\text{p},k}^{(n)}$ and $\M{C}_{\text{p},k}^{(n)}$ denote the mean and covariance matrix of the prediction message $\phi^{\text{p}}_{\rightarrow n}(\V{p}_{k}^{(n)})$ related to position $\V p_k ^{(n)}$, respectively.
\item Second, we perform a Bayesian update step with the mean $\bar {\V{\lambda}}_j^{(n)}$ and covariance matrix $\overline{ \M {\Sigma}}_j^{(n)}$ as an input by using sigma points to approximate the nonlinear measurement model described by the ``likelihood'' $f(\overline{\V{z}}_{j}^{(n)}|\overline{\V{x}}_{j}^{(n)})$ (see \cite{WanMer:B01} for details). This results in an approximate updated mean $\overline{\V{\mu}}_{j}^{(n)}$ and  covariance matrix $\overline{\M{C}}_{j}^{(n)}$ representing the ``stacked belief'' $b(\overline{\V{x}}_{j}^{(n)})$ in \eqref{eq:bp_hd}.
\item Third, we perform a marginalization step by extracting the mean $\V{\mu}_{j}^{(n)}$ and the covariance matrix $\M{C}_{j}^{(n)}$ from $\overline{\V{\mu}}_{j}^{(n)}$ and $\overline{\M{C}}_{j}^{(n)}$, respectively, as a representation of $b(\V{x}_{j}^{(n)})$. Specifically, $\V{\mu}^{(n)}_{j}$ is given by the first $N_x$ elements of $\overline{\V{\mu}}^{(n)}_{j}$, and $\M{C}^{(n)}_{j}$ is given by the upper-left $N_x \ist\times\ist N_x$ submatrix of $\overline{\M{C}}^{(n)}_{j}$.%
\end{enumerate}
After the $\V{\mu}_{j}^{(n)}$ and $\M{C}_{j}^{(n)}$ are computed with the \ac{spbp} algorithm, the \ac{mmse} estimate $\hat {\V x}_j^{(n)}$ can be interpreted as $\V{\mu}_{j}^{(n)}$.
The node inference algorithm is summarized in \cite{MeyEtzLiuHlaWin:J18}.

\begin{remark}
In the Peregrine implementation, every node $j$ can compute the message passing equations \eqref{eq:bp_marg}--\eqref{eq:bp_meas} even in the absence of a synchronized network. This is because, as will be discussed in Section \ref{sec:implementation}, means $\V{\mu}_{\text p,k}^{(n)}$ and covariances $\M{C}_{k}^{(n)}$ \vspace{-.25mm} related to the prediction message $\phi_{\rightarrow n}(\V{x}_{k}^{(n)}) $ of the measured nodes $k \in \Nnb{j}^{(n)}$ are received while range measurements are performed. Furthermore, in case no measurements are performed by agent $j \in \Nagt$, its belief can directly be obtained from \eqref{eq:bp_marg} and \eqref{eq:bp_pred} as $b(\V{x}_{j}^{(n)}) = \phi_{\rightarrow n}(\V{x}_{j}^{(n)})$.
\end{remark}

\begin{remark} Even though vectors with augmented dimensions are used in the \ac{spbp} algorithm, the computational complexity of node inference is still low. In particular, the main computational complexity of the node inference algorithm in \cite{MeyEtzLiuHlaWin:J18} results from the calculation of the mean and the covariance matrix representing $b(\overline{\V{x}}_{j}^{(n)})$ in the measurement update step. The complexity of this operation is cubic with respect to the dimension of $\overline{\V{x}}_{j}^{(n)}$ \cite{JulUhl:C97}. In 3-D scenarios, the dimension of $\overline{\V{x}}_{j}^{(n)}$ is $3 |\Set{M}^{(n)}_{j}| + N_x$. This means that the complexity depends on the number of non-zero elements of the measurement control vector $\V{m}_{j}^{(n)}$ and can thus be controlled by the node prioritization algorithm described in Section \ref{sec:nodePrioritization}.
\end{remark}
\section{Network Operation}
\label{sec:operation}

In this section, we first present preliminaries on the agent position uncertainty, which is used for evaluating the localization performance of our network operation algorithms. We then present the node activation and node prioritization algorithms used to control network operations.

\subsection{Preliminaries}
\label{sec:preSPEB}
As the density of the wireless network increases, the contention for wireless channels becomes stronger and the available communication resources for each agent become more limited. Our network operation strategy improve the scalability of Peregrine by carefully managing the contention for wireless channels and efficiently utilizing the communication resources in the network. In particular, Peregrine integrates the node activation strategy \cite{WanSheConWin:J17} with node prioritization strategy \cite{DaiSheWin:J15a} presented in our previous work. Such integration is nontrivial. First, existing node activation and node prioritization algorithms have been designed independently. In contrast, we use the measurement control vector provided by the node prioritization strategy as an input for the node activation strategy, thereby integrating the two strategies in a systematic manner. Second, existing algorithms do not or only partially consider the limitations of actual hardware (e.g., the limited dynamic range of radio receivers). These limitations are taken into fully consideration in Peregrine.

We use the trace of the covariance matrix of the position estimate as the metric to characterizes uncertainty in the network operation algorithms. The reason for not using the \ac{mse} is that it is intractable in real time because of the unavailability of the true node positions. The covariance matrix of the position estimate, which we refer to as \textit{position covariance matrix} for short, can be computed based on the posterior distribution of the agent's state. However, it is infeasible to evaluate the exact position covariance matrix in general, and therefore we compute their approximate values instead. In particular, consider the evolution of the position covariance matrix of agent $j$ at time $n$ after it makes $m_{jk}^{(n)}$ distance measurements with neighbor $k \in \Set M_j^{(n)}$. To compute the updated position covariance matrix $\breve {\M{C}}\big(\V m_j^{(n)}\big)$ of agent $j$ after performing these measurements, we make the following approximations: first, the joint distribution of $[\RV p_{j'}^{(n)}]_{j' \in \{j\}\cup \Set M_{j}^{(n)}}$ \textit{before} node $j$ makes measurements is approximated by Gaussian distribution $\mathcal N(\overline{\V \lambda}_{\mathrm p, j}^{(n)}, \overline{\M \Sigma}_{\mathrm p, j}^{(n)})$, where $\overline{\V \lambda}_{\mathrm p, j}^{(n)}$ and $\overline{\M \Sigma}_{\mathrm p, j}^{(n)}$ are defined as
\begin{align}
\overline{\V \lambda}_{\mathrm p, j}^{(n)} &\triangleq \big[ {\V{\mu}}_{\mathrm p, j}^{(n) \ist \text{T}} \;\, \V{\mu}_{\text{p},k_1}^{(n) \ist \text{T}} \;\,  \V{\mu}_{\text{p},k_2}^{(n) \ist \text{T}} \cdots\ist \V{\mu}_{\text{p}, k_{|\Set{M}^{(n)}_j|}}^{(n) \ist \text{T}} \big]^{\text{T}} \rmv \nn\\[1mm]
\overline{\M {\Sigma}}_{\mathrm p, j}^{(n)} &\triangleq \mathrm{bdiag} \Big\{ {\M{C}}_{\mathrm p, j}^{(n)} ~ \M{C}_{\text{p},k_1}^{(n)} ~ \M{C}_{\text{p},k_2}^{(n)} \cdots ~ \M{C}_{\text{p},k_{|\Set{M}^{(n)}_j|}}^{(n)} \Big\}\, . \label{eq:stackedPos}
\end{align}
Comparing \eqref{eq:stackedPos} with \eqref{eq:stackedMeanCov}, we note that this is a natural approximation based on the \ac{spbp} algorithm in Section~\ref{sec:inference}. Second, we approximate the measurement model \eqref{eq:meas_def} via local linearization. Specifically, \eqref{eq:meas_def} is a nonlinear function of agent positions $\big[\RV p_j^{(n) \mathrm T} ~ \RV p_k^{(n) \mathrm T}\big]^{\mathrm T}$, and we approximate it by its first-order Taylor expansion at $\big[{\V \mu}_{\text p, j}^{(n) \text T}  ~ \V \mu_{\text p, k}^{(n) \text T}  \big]^{\text T}$. Such approximation can be simplified as
\begin{align}\label{eq:meas_app}
	\rv{z}_{jk}^{(n)} \approx \udv_{jk}^{(n) \mathrm T} \big (\RV{p}_{k}^{(n)} - \RV{p}_{j}^{(n)} \big) + \rv{q}_{jk}^{(n)}
\end{align}
where $\udv_{jk}^{(n)}$ is the estimate of the unit direction vector between nodes $j$ and $k$ given by
\begin{align} \label{eq:udv}
\udv_{jk}^{(n)} =  \frac{{\V{\mu}}_{\text p,k}^{(n)} - {\V{\mu}}_{\text p, j}^{(n)}  }   {\big \| {\V{\mu}}_{\text p,k}^{(n)} - {\V{\mu}}_{\text p, j}^{(n)} \big \|} \,.
\end{align}
The second approximation has been widely adopted in the literature \cite{AruMasGorCla:02} for dealing with nonlinear measurement models. Based on these two approximations, the posterior distribution of $[\RV p_{j'}^{(n)}]_{j' \in \{j\}\cup \Set M_{j}^{(n)}}$ \textit{after} node $j$ makes measurements $\rv z_{jk}^{(n)}, k \in \Set M_{j}^{(n)}$ is Gaussian. Moreover, the covariance matrix $\breve {\M{C}}\big(\V m_j^{(n)}\big)$ for $\RV p_j^{(n)}$ in such distribution is
\begin{IEEEeqnarray}{C}\label{eq:na_fim}
\breve {\M{C}}\big(\V m_j^{(n)}\big) = \Big [ \big[{\M C}_{\text p,j}^{(n)}\big] ^{-1}   \rmv \rmv \rmv + \rmv \rmv \rmv \rmv \sum_{k\in \Set M_{j}^{(n)}} \rmv \rmv \rmv \rmv \rmv \cerc_{jk}^{(n)}\big(m_{jk}^{(n)}\big)  \V u_{jk}^{(n)}   \V u_{jk}^{(n) \text T}  \Big]^{-1} \IEEEeqnarraynumspace
\end{IEEEeqnarray}
where $\cerc_{jk}^{(n)}\big(m_{jk}^{(n)} \big) $ is the intensity of information agent $j$ obtains from the $m_{jk}^{(n)}$ distance measurements with node $k$, and it is given by
\begin{align}
\label{eq:cerc1}
\cerc_{jk}^{(n)}\big(m_{jk}^{(n)}\big)  = \frac{m_{jk}^{(n)} \erc_{jk}^{(n)}} {1 + m_{jk}^{(n)} \erc_{jk}^{(n)}  \V u_{jk}^{(n) \text T}   \M{C}_{\text{p},k}^{(n)} \V u_{jk}^{(n)} }.
\end{align}
Note that  the position uncertainty of agent $k$ is taken into account in evaluating $\cerc_{jk}^{(n)}\big(m_{jk}^{(n)} \big)$ according to \eqref{eq:cerc1}. On one hand, if $k \in \Nagt$ and $\M{C}_{\text{p},k}^{(n)}$ is positive definite, then $\cerc_{jk}^{(n)}\big(m_{jk}^{(n)} \big) \leq \big [ \V u_{jk}^{(n) \text T}  \M{C}_{\text{p},k}^{(n)} \V u_{jk}^{(n)} \big ]^{-1}$, $\forall m_{jk}^{(n)} \geq 0$. In other words, the information node $j$ can obtain from the distance measurements with node $k$ is limited because of the uncertainty of $\RV p_k^{(n)}$. On the other hand, if $k \in \Nanc$ we have $ \M{C}_{\text{p},k}^{(n)} = \M 0_{3 \times 3}$, which according to \eqref{eq:cerc1} results in $\cerc\big(m_{jk}^{(n)}\big)  = m_{jk}^{(n)}\erc_{jk}^{(n)}$. 

A key observation from \eqref{eq:na_fim} is that the evolution of the position covariance matrix after node $j$ makes distance measurements with its neighbors at time $n$ is the same as that of the inverse \ac{efim} \cite{SheWymWin:J10, DaiSheWin:J15, WanSheConWin:J17}. Since the network operation algorithms presented in \cite{SheWymWin:J10, DaiSheWin:J15, WanSheConWin:J17} use the trace of the inverse \ac{efim} as the performance metric, we can adapt them for Peregrine by replacing inverse \acp{efim} with position covariance matrices.

\subsection{Node Activation}
\label{sec:nodeActivation}

Multiple agents need to share a common channel, especially as the density of the wireless network increases. To avoid communication interference, only a subset of agents can be activated to access the channel and make distance measurements with their neighbors in a certain short time window. The goal of node activation is to determine in a distributed manner which agents should access the channel to minimize communication interference and to improve the overall localization accuracy in the network. 

We adopt the node activation strategy from \cite{WanSheConWin:J17}, hereinafter referred to as \ac{htna}. An agent first senses the channel for a random period of time. If the channel is idle during that period, the agent evaluates the potential improvement in its own localization accuracy related to accessing the channel. A specific improvement can be calculated from the proposed measurement vector $\V{m}^{(n)}_j$ provided by node prioritization. If the localization performance improvement is significantly high, the agent attempts to access the channel; otherwise it remains silent to give other agents the opportunity to make measurements and improve their localization accuracy.

In particular, consider the node activation process for agent $j \in \Nagt$ at time $n$. Let binary variable $\cai_j^{(n)}$ denote the channel access indicator such that $j$ will attempt to access the channel if $\cai_j^{(n)} = 1$ and it will remain silent  otherwise. The expression for $\cai_j^{(n)}$ is given by
\begin{align}\label{eq:activation_thresh}
	{\cai}_j^{(n)} = \begin{cases}
		1, & \quad \text{if}\ \eer_{j}^{(n)}\big(\V m_j^{(n)}\big) > \eei_{{j}}^{(n)} \\
		0, & \quad \text{otherwise},
	\end{cases} 
\end{align}
where $\eer_{j}^{(n)}\big(\V m_j^{(n)}\big)$ is the potential \textit{trace reduction of the position covariance matrix} of agent $j$ from the distance measurements it makes according to the control vector $\V m_j^{(n)}$, and $\eei_{{j}}^{(n)}$ is the total \textit{trace increase of the position covariance matrix}  of the subnetwork $\{j\} \cup \Nnb{j}^{(n)} $ while channel access is performed in order to measure the distance to each neighbor $k \in \Set{M}_j^{(n)}$, $m^{(n)}_{jk}$ times (We recall that $\Nnb{j}^{(n)}$ is the set of nodes that are able to make distance measurements with agent $j$ at time step $n$, whereas $\Set{M}_j^{(n)}$ consists of $k \in \Nnb{j}^{(n)}$ such that $m^{(n)}_{jk} > 0$). %
Specifically, the trace reduction $\eer_{j}^{(n)}\big(\V m_j^{(n)}\big)$ is
\begin{align*}
\eer_{j}^{(n)}\big(\V m_j^{(n)}\big) = \tr \big\{{\M{C}}_{\text p,j}^{(n)} \big\} - \tr \big\{ \breve {\M{C}}\big(\V m_j^{(n)}\big) \big \} 
\end{align*}
where we recall $\breve {\M{C}}\big(\V m_j^{(n)}\big)$ is given in \eqref{eq:na_fim}. The trace increase $\eei_{{j}}^{(n)}$ is derived according to the state-evolution model \eqref{eq:cvm}. Specifically, $\eei_{{j}}^{(n)}$ is given by
\begin{align} \label{eq:error_increase_tot}
	\eei_{{j}}^{(n)} = \hspace{-6mm} \sum_{k \in \{ j \} \cup \Nnb{j}^{(n)} \setminus \Nanc}   \hspace{-6mm}  \tr\Big\{ \eim_{\text{p},k}^{(n)} \Big\}
	\vspace{-1.5mm}
\end{align}
where $\eim_{\text{p},k}^{(n)}$ is the position-related $3 \times 3$ upper-left submatrix of the covariance increase matrix $\eim_{k}^{(n)}$ given by
\begin{align} \label{eq:errorIncrease}
	\eim_{k}^{(n)} = \M{A}({t^{(n)}_{j}})\M{C}_{k}^{(n)} \rmv\rmv \M{A}^{\text{T}} ({t^{(n)}_{j}})+  \M{C}_w(t^{(n)}) - \M{C}_{k}^{(n)} \rmv\rmv.
\end{align}
Furthermore, $t^{(n)}_j$ is the assumed channel access time related to performing the measurements with neighbors $k \in \Set{M}_j^{(n)}$ as controlled by $\V{m}^{(n)}_j$.
\vspace{1mm}

\begin{remark}
Note that node activation is performed in an asynchronous way by agents in the network. For notational simplicity, we denote  the most recent versions locally available at a specific agent with index $n$ in this section. For example,  $\M{C}_{k}^{(n)}$ in  \eqref{eq:errorIncrease} denotes the most recent covariance matrix of node $k \in \Set{N}_j^{(n)}$ that was transmitted to agent \vspace{.3mm} $j$.
\end{remark}
The \ac{htna} algorithm has the following favorable properties:
\begin{itemize}
\item  It adapts to the network size. Consider adding an agent to the sub-network formed by $ \{j\} \cup \Nnb{j}^{(n)} $. On one hand, more agents will contend for the channel access. On the other hand, the value $\delta_{{j}}^{(n)}$ also increases according to \eqref{eq:error_increase_tot} for each agent $k \in \{j\} \cup \Nnb{j}^{(n)}$, and thus the chance that $\cai_k^{(n)} = 1$ for each single agent $k \in \{j\} \cup \Nnb{j}^{(n)}$ decreases. As a result, a smaller subset of nodes in the sub-network who can benefit the most from the distance measurements will actually attempt to access the channel, and the possibility that two agents try to access the channel at the same time is reduced.
\vspace{1.5mm}

\item The computation complexity and communication overhead for evaluating $\cai_j^{(n)}$ is small. In particular, the complexity for computing $\eer_j^{(n)}\big(\V m_j^{(n) }\big)$ and $\eei_j^{(n)}$  grows only linearly with the number of neighbors of $j$. Moreover, each agent $k \in \Nnb{j}^{(n)}$ only needs to send an $N_x \times N_x$ matrix $\M{C}_{k}^{(n)}$ and a $3 \times 1$ vector $\V{\mu}_{\text p,k}^{(n)}$ to $j$. 
\vspace{-.2mm}
\end{itemize}
Details of the \ac{htna} algorithm for agent $j$ at time $n$ are shown in Algorithm~\ref{alg:node_activation}: lines 3--7 compute the trace reduction of the position covariance matrix; lines 8--12 compute the trace increase of the position covariance matrix.
\begin{algorithm}[t]
\caption{Holistic Threshold-Based Node Activation}\label{alg:node_activation}
\begin{algorithmic}[1]
\Require $\V{\mu}_{\text p, j}^{(n)}$, $\M{C}_{\mathrm{p},j}^{(n)}$;  $\V \mu_{\text p, k}^{(n)}$, $\M C_{k}^{(n)}$ for $k \in \Nnb{j}^{(n)}$; $\V m_j^{(n)}$.
\Ensure $\{z_{jk}^{(n)}, k \in \Set M_{j}^{(n)}\}$;

	\State Sense the channel for random amount of time;
	\If {channel stayed idle in the sensing period}
		\ForAll{$k \in \Set M_j^{(n)}$}
			\State %
			$\begin{aligned}
			\udv_{jk}^{(n)} & \gets \frac{\V{\mu}_{\text p,k}^{(n)} - \V{\mu}_{\text p, j}^{(n)}} {  \big \| \V{\mu}_{\text p,k}^{(n)} - \V{\mu}_{\text p, j}^{(n)} \big \|} \\
			\cerc_{jk}^{(n)}(m_{jk}^{(n)}) & \gets \frac{m_{jk}^{(n)}\erc_{jk}^{(n)} } {1 + m_{jk}^{(n)} \erc_{jk}^{(n)} \V u_{jk}^{(n) \text T}  \M{C}_{\text{p},k}^{(n)} \V u_{jk}^{(n)}  }
			\end{aligned}$
		\EndFor
		\State Compute $\breve {\M{C}}\big(\V m_j^{(n)}\big)$ according to \eqref{eq:na_fim}
		\State
		$\begin{aligned}
		\eer_{j}^{(n)}\big(\V m_j^{(n)}\big) \gets \tr \big\{\M{C}_{\text p,j}^{(n)} \big\} - \tr \big\{ \breve {\M{C}}\big(\V m_j^{(n)}\big) \big \}
		\end{aligned}$
		\ForAll{$k \in \Nnb{j}^{(n)} \cup \{j\} \setminus \Nanc$}
			\State %
			$\begin{aligned} 
			\eim_{k}^{(n)} \gets & \M{A}(t^{(n)}_{j})\M{C}_{k}^{(n)} \M{A}^{\text{T}} (t^{(n)}_{j}) + \M{C}_w(t^{(n)}) - \M{C}_{k}^{(n)} 
			\end{aligned}$
			\State \parbox[t]{\dimexpr\linewidth- \algorithmicindent * 2}{Obtain $\eim_{\text{p},k}^{(n)}$ by extracting the $3 \times 3$ upper-left submatrix of $\eim_{k}^{(n)}$;\strut}              
		\EndFor	
		\State 
		$\begin{aligned}
		\eei_{{j}}^{(n)} \gets \hspace{-6mm} \sum_{k \in \Nnb{j}^{(n)} \cup \{ j \}\setminus \Nanc} \hspace{-6mm} \tr\Big\{ \eim_{\text{p},k}^{(n)} \Big\}
		\end{aligned}$
		\If {$\eer_j^{(n)}\big(\V m_j^{(n)} \big)> \eei_j^{(n)}$}
			\State \parbox[t]{\dimexpr\linewidth- \algorithmicindent * 2} {Perform measurements $z_{jk}^{(n)}$ with all $k \in \Set{M}^{(n)}_j$ as defined by $\V{m}^{(n)}_{j}$.\strut} 
		\Else
			\State $\{z_{jk}^{(n)}, k \in \Set M_{j}^{(n)}\} \gets \varnothing$
		\EndIf
	\EndIf
\end{algorithmic}
\end{algorithm}

\subsection{Node Prioritization}
\label{sec:nodePrioritization}

Each agent may have many neighbors in its local subnetwork and therefore many possible range measurements. Yet, a network has finite communication resources that agents can use to make measurements. In this discussion, we consider the resource to be time or equivalently, the number of measurements. Specifically, the nodes contend for the time they need to make measurements on a shared channel. An equivalent discussion could focus on allocating bandwidth or transmit power as they are examples of the more general energy allocation problem. The goal of node prioritization is to devote these finite resources to the most beneficial measurements.

For node prioritization, we adapt the strategy in \cite{DaiSheWin:J15a}, hereinafter referred to as \ac{cpnp}. At each time step, an agent determines the measurement allocation scheme by solving an optimization problem. The objective of this optimization problem is to minimize the trace of position covariance matrix of each agent given a constraint on the total number of distance measurements the agent can make at a time step. The solution to this problem is a control vector containing the number of distance measurements the agent should make with each of its neighbors. This vector of ``proposed measurements'' is used as an input for the node activation algorithm.

Consider the node prioritization process for agent $j \in \Nagt$ at time $n$ given the constraint that it can make no more than $M_j^{(n)}$ measurements in total with all its neighbors. Under such constraint, agent $j$ determines the number of measurements it makes with each of its neighbors that minimizes the trace of its own position covariance matrix. This task is equivalent to solving the optimization problem $\mathscr{P}_{j}^{(n)}$ expressed as
\begin{align}
	\mathscr{P}_{j}^{(n)}:
    \min_{\{\tilde{\V{m}}^{(n)}_{j}\}} & \quad \tr \big \{  \breve {\M{C}}\big(\tilde{\V m}_j^{(n)}\big)  \big \} \nonumber \\
	\text{s.t.} & \quad \tilde{m}^{(n)}_{jk} \in \mathbb N_0, \qquad \forall k \in \Nnb{j}^{(n)} \label{eq:np_int} \\
	& \quad \sum_{k \in \Nnb{j}^{(n)}} \tilde{m}^{(n)}_{jk} \leq M_{j}^{(n)} \label{eq:np_tot}
\end{align}
where $\tilde {\V{m}}^{(n)}_j = [\tilde{m}^{(n)}_{jk}]_{k \in \Nnb{j}^{(n)}}$ is the optimization variable of the problem with element $\tilde{m}^{(n)}_{jk}$ representing the number of distance measurements allocated to neighbor $k$; $ \breve {\M{C}}\big(\tilde{\V m}_j^{(n)}\big) $ is the position covariance matrix after the distance measurements indicated by $\tilde {\V m}_j^{(n)}$ are made, and it is given in \eqref{eq:na_fim} by replacing ${\V{m}}^{(n)}_j$  with $\tilde{\V{m}}^{(n)}_j$; $\mathbb N_0$ represents the set of non-negative integers. 

The problem $\mathscr{P}_{j}^{(n)}$ involves integer constraints \eqref{eq:np_int} and is thus hard to solve. Therefore, we solve a relaxed problem $\mathscr{P}_{\text R, j}^{(n)}$ instead by replacing \eqref{eq:np_int} with nonnegative constraints $\tilde{m}^{(n)}_{jk} \geq 0$. Using the same method as presented in \cite{DaiSheWin:J15a}, we show that this relaxed problem can be transformed to a \ac{sdp} problem given by
\begin{align}
	\mathscr{P}_{\text R, j}^{(n)}: \rmv \rmv \rmv \rmv
    \min_{\{\tilde{ \V{m}}^{(n)}_{j}, \V y_j, \M{M} \}} & \quad \tr \{ \M M\} \nonumber \\  
	\text{s.t.} & \quad 
				\begin{bmatrix}
				\M M & \M I_3 \\
				\M I_3 & \M J(\V y_j)
				\end{bmatrix} \succeq \M 0_{6 \times 6} \nonumber \\
			& \quad  y_{jk} \leq \tilde {{m}}^{(n)}_{jk}, \quad k \in \Set{N}^{(n)}_j \nonumber \\
			& \quad y_{jk} \geq 0, \qquad \hspace{2mm}k \in \Set{N}^{(n)}_j \nonumber  \\[.8mm]
			& \quad \big \| \big[ \sqrt{2}  \hspace{3mm}  1\rmv-\rmv \rho_{jk}^{(n)} y_{jk}  \hspace{3mm} 1\rmv+\rmv \rho_{jk}^{(n)} \tilde m_{jk}^{(n)}  \big ]^{\text T} \big \| \nonumber \\[.8mm]
			& \qquad \leq \rho_{jk}^{(n)} \big(\tilde m_{jk}^{(n)} - y_{jk}\big)  + 2  \label{eq:np_constraint_socp} \\[.8mm]
			& \quad  \sum_{k \in \Nnb{j}^{(n)}} \tilde{m}^{(n)}_{jk} \leq M_{j}^{(n)}, \label{eq:np_prelax} \\[-7mm]
			\nonumber
\end{align}
where we introduced the short notation $\rho_{jk}^{(n)} = \erc_{jk}^{(n)}  \V u_{jk}^{(n)\text T}  \M{C}_{\text{p},k}^{(n)} \V u_{jk}^{(n)} $; matrix $\M J(\V y_j)$ is a short notation given by
\begin{align*}
\M J(\V y_j) =  \big[\M{C}_{\text p,j}^{(n)}\big] ^{-1}  \rmv + \rmv \sum_{k\in \Set M_{j}^{(n)}} y_{jk} \erc_{jk}^{(n)}  \V u_{jk}^{(n)}   \V u_{jk}^{(n) \text T};
\end{align*} 
$\V y_j = [y_{jk}]_{k \in \Nnb{j}^{(n)}}$ and matrix $\M M$ are auxiliary optimization variables introduced for converting the relaxed problem to an \ac{sdp} problem;
The size of $\mathscr{P}_{\text R, j}^{(n)}$ is as follows: the total number of optimization variables is $2 | \Set{N}^{(n)}_j| + 6$; the total dimension $n_{\mathrm c}$ of the \ac{lmi} constraints is $n_{\mathrm c} = 6 | \Set{N}^{(n)}_j| + 7$, as \eqref{eq:np_constraint_socp} can be converted to an \ac{lmi} of dimension $4$ for each $k \in \Set{N}^{(n)}_j$ \cite{VanBoy:96}. If interior-points methods are adopted, the worst-case number of iterations required for solving the \ac{sdp} $\mathscr{P}_{\text R, j}^{(n)}$ increases as $O\big(n_{\mathrm c}^{1/2}\big)$, whereas fewer number of iterations are actually required in practice \cite{VanBoy:96}.%
The solution $\{ \V m_j^{(n)\hspace{.1mm}*}\rmv\rmv\rmv\rmv, \V y_j^*, \M M^* \}$ to problem $\mathscr{P}_{\text R, j}^{(n)}$ is obtained using a convex optimization engine and $\V m_j^{(n)\hspace{.1mm}*}$ contains non-integer components in general \cite{DiaBoy:cvxpy}. We round the non-integer components to integer values to obtain a solution $\V m_j^{(n)}$ to the original problem $\mathscr{P}_{j}^{(n)}$.

Different from \cite{DaiSheWin:J15a}, we search for a method to allocate the number of distance measurements instead of the transmit power to each neighbor. In \cite{DaiSheWin:J15a}, the objective function depends on the received signal energy from each neighbor, and it is assumed in \cite{DaiSheWin:J15a} that such energy has a linear relationship with the transmit power.
In practical systems, radio receivers have fixed instantaneous dynamic ranges, often smaller than the dynamic range needed for the maximum and minimum expected separation of the radios. In such situations radios use automatic gain control to maintain a high {stable} {signal-to-noise-ratio} in the receiver. As a result, it is not straightforward to characterize the relation between the receive signal energy and the transmit power. Therefore, we fix the transmit power for an individual distance measurement, and optimize over the number of measurements to make in the \ac{cpnp} algorithm. In this case, the receive signal energy from a neighbor has a linear relationship with the number of measurements.
\begin{algorithm}[t]
\caption{Conic Programming-based Node Prioritization}\label{alg:node_prioritization}
\begin{algorithmic}[1]
\Require $\V{\mu}_{\text p, j}^{(n)}$, $\M{C}_{\mathrm{p},j}^{(n)}$;  $\V \mu_{\text p, k}^{(n)}$, $\M C_{\text p, k}^{(n)}$ for $k \in \Nnb{j}^{(n)}$; \ac{erc} $\erc_{jk}^{(n)}$ for $k \in \Nnb{j}^{(n)}$.
\Ensure Control vector $\V m_j^{(n)} = [m_{jk}^{(n)}]_{k \in \Nnb{j}^{(n)}}$;
	\ForAll{$k \in \Nnb j^{(n)}$}
		\State %
		$\begin{aligned}
		\udv_{jk}^{(n)} & \gets \frac{\V{\mu}_{\text p,k}^{(n)} - \V{\mu}_{\text p, j}^{(n)}} { \big \| \V{\mu}_{\text p,k}^{(n)} - \V{\mu}_{\text p, j}^{(n)} \big \| }
		\end{aligned}$
	\EndFor

	\State Compute the solution $\{\V m_j^{*\ist(n)}, \V y_j^{*},  \M M^{*}\}$ to problem $\mathscr{P}_{\text R, j}^{(n)}$ given by \eqref{eq:np_prelax}
	\ForAll{$k \in \Nnb j^{(n)}$}
		\State	$m_{jk}^{(n)} \gets \mathrm{round}( m_{jk}^{*(n)})$
	\EndFor
\end{algorithmic}
\end{algorithm}
The \ac{cpnp} algorithm does not introduce extra communication overhead. In particular, in the \ac{cpnp} algorithm, agent $j$ requires $\V \mu_{\text p, k}^{(n)}$ and $\M C_{\text p, k}^{(n)}$ for all neighbors $k \in \Nnb{j}^{(n)}$, which are already used by the node activation and node inference algorithms. The \ac{cpnp} algorithm is summarized in Algorithm \ref{alg:node_prioritization} for agent $j$ and a certain time $n$, where $\mathrm{round}(\cdot)$ is a scalar function that rounds its argument to the closest integer.
\section{System Implementation} 
\label{sec:implementation}

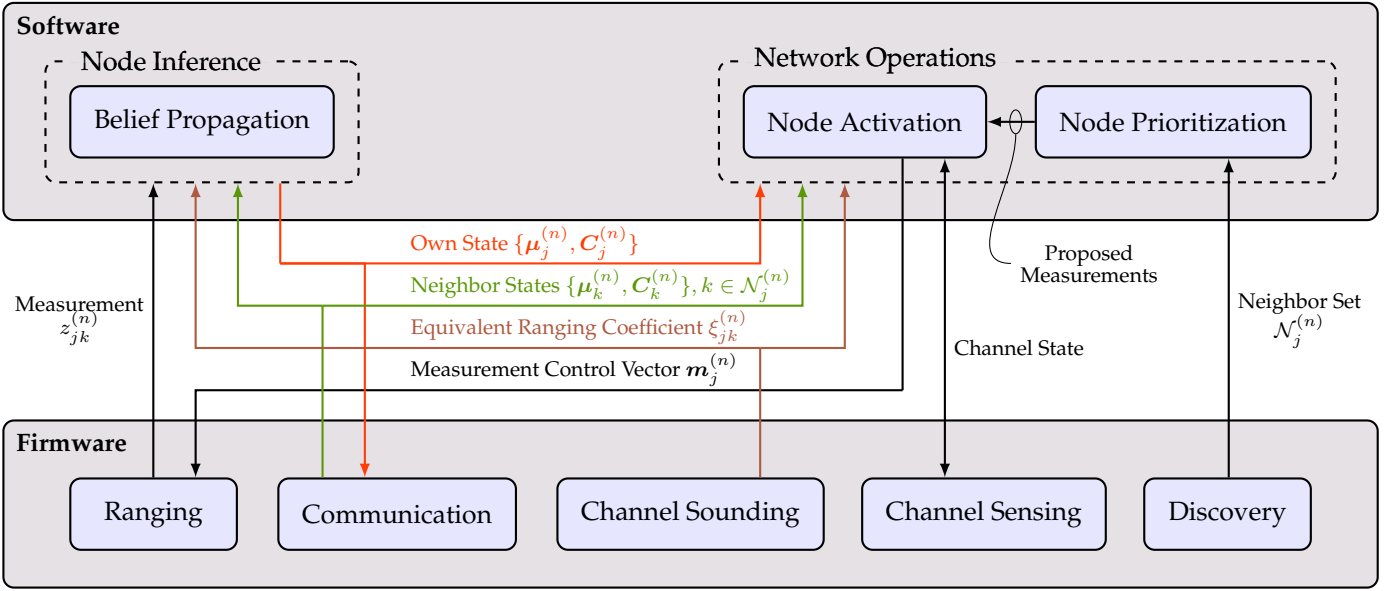
\begin{figure*}[t]
	\begin{minipage}[c]{\textwidth}
	\centering
	\begin{tikzpicture}

	\def\levelwidth{510pt}


	\node[block] (spbp) {Belief Propagation};
	\mkfunction{ni}{(spbp)}{Node Inference};
	
	\node[block, right=470pt of spbp.west, anchor=east] (np) {Node Prioritization};

	\node[block, left=of np.west] (na) {Node Activation};
	\mkfunction{no}{(na)(np)}{Network Operations};
	\mklevel[minimum width=\levelwidth]{sw}{(spbp)(ni)(np)(na)(no)}{Software};

	\def\levelsep{148.5pt}
	\node[block, below=\levelsep of spbp.west, anchor=west] (rng) {Ranging};

	\node[block, below=\levelsep of np.east, anchor=east] (disc) {Discovery};
	\path (rng) -- (disc) node[midway, block] (csound) {Channel Sounding};
	\path (rng) -- (csound) node[midway, block] (comm) {Communication};

	\path (csound) -- (disc) node[midway, block] (csens) {Channel Sensing};
	\mklevel[minimum width=\levelwidth]{fw}{(rng)(disc)(csound)(comm)(csens)}{Firmware};

	\def\aspace{16pt}
	\def\haspace{(\aspace, 0pt)}
	\def\mhaspace{(-\aspace, 0pt)}
	\def\vaspace{(0pt, \aspace)}
	\def\mvaspace{(0pt, -\aspace)}
	\path (sw.south) coordinate (bus_0)
		++\mvaspace coordinate coordinate (bus_1)
		++\mvaspace coordinate coordinate (bus_2)
		++\mvaspace coordinate coordinate (bus_3)
		++\mvaspace coordinate coordinate (bus_4)
		($(sw.south)!0.5!(fw.north)$) coordinate (bus_mid);

	\path (ni.south east) coordinate (ni_a0)
		++\haspace coordinate (dummy)
		++(4pt, 0pt) coordinate (text_align);

	\path (rng.north east) coordinate (rng_a0)
		++\mhaspace coordinate (rng_a2)
		++\mhaspace coordinate (rng_a1);

	\path (rng_a1 |- ni.south) coordinate (ni_a1)
		++\haspace coordinate (ni_a2)
		++\haspace coordinate (ni_a3)
		++\haspace coordinate (ni_a4);
	\path (no.south west) coordinate (no_a0)
		++\haspace coordinate (no_a1)
		++\haspace coordinate (no_a2)
		++\haspace coordinate (no_a3);
	\path (na.south east) coordinate (na_a0)
		++\mhaspace coordinate (na_a2)
		++\mhaspace coordinate (na_a1);
	\path ($(np.south west)!0.7!(np.south east)$) coordinate (np_a1);
	
	\path (ni_a4 |- comm.north) coordinate (comm_a0)
		++\haspace coordinate (comm_a1)
		++\haspace coordinate (comm_a2)
		++\haspace coordinate (text_align);
	\path (no_a1 |- csound.north) coordinate (csound_a1);

	\draw[-latex] (rng_a1 -| ni_a1) -- (ni_a1 |- bus_mid) node[left, align=center] {Measurement \\ $z_{jk}^{(n)}$} -- (ni_a1); 
	\draw (na_a1) -- (na_a1 |- bus_4) coordinate (mc_1);
	\draw[-latex] (mc_1) -- (text_align |- mc_1) node[above right, inner sep=1pt] {Measurement Control Vector $\V{m}_{j}^{(n)}$} -| (rng_a2); 
	\draw[pastel9] (ni_a4) -- (ni_a4 |- bus_1) coordinate (os_1);
	\draw[-latex, pastel9] (os_1) -| (comm_a2);
	\draw[-latex, pastel9] (os_1) -- (text_align |- os_1) node[above right, inner sep=1pt] {Own State $\{\V{\mu}_{j}^{(n)}, \M{C}_{j}^{(n)}\}$} -| (no_a1); 
	\draw[pastel3] (comm_a1) -- (comm_a1 |- bus_2) coordinate (ns_1);
	\draw[-latex, pastel3] (ns_1) -| (ni_a3);
	\draw[-latex, pastel3] (ns_1) -- (text_align |- ns_1) node[above right, inner sep=1pt] {Neighbor States $\{\V{\mu}_{k}^{(n)}, \M{C}_{k}^{(n)}\}, k \in \Set{N}_j^{(n)}$} -| (no_a2); 
	\draw[pastel4] (csound_a1) -- (csound_a1 |- bus_3) coordinate (erc_1);
	\draw[-latex, pastel4] (erc_1) -- (text_align |- erc_1) node[above right, inner sep=1pt] {Equivalent Ranging Coefficient $\xi^{(n)}_{jk}$} -| (ni_a2);
	\draw[-latex, pastel4] (erc_1) -| (no_a3); 
	\draw[latex-latex] (na_a2 |- csens.north) -- (na_a2 |- bus_3) node[ right, align=center] {Channel State} -- (na_a2); 
	\draw[-latex] (disc.north -| np_a1) -- (np_a1 |- bus_mid) node[right, align=center] {Neighbor Set \\ $\Set{N}_j^{(n)}$} -- (np_a1); 
	\draw[-latex] (np) -- (na) coordinate[midway] (pm);
	\node[right, align=center] (pm_label) at (bus_1 -| pm) {Proposed \\ Measurements};
	\node[ellipse, thin, draw=black, minimum height=9pt, minimum width=4.5pt, inner sep=0pt] at ($(np.west)!0.4!(na.east)$) (pm_ell) {};
	\draw[thin] (pm_label.west) to [out=180, in=270] (pm_ell.south);

\end{tikzpicture}
	\caption[Software/firmware architecture block diagram]{Block diagram of the software/firmware architecture running on a single Peregrine node $j$.}
	\label{fig:architecture}
	\end{minipage}
\end{figure*}

Peregrine is implemented as a distributed system involving three major components: the \textit{software} consisting of the \ac{nln} algorithms described in Sections~\ref{sec:inference} and \ref{sec:operation}; the \textit{firmware} which includes our cooperative ranging protocol; and the \textit{hardware}, a number of identical battery-powered sensing and processing hardware nodes. A block diagram of the software and firmware is shown in Fig.~\ref{fig:architecture}.

\subsection{Software} 
\label{sub:software}

The \ac{nln} algorithms described in Sections~\ref{sec:inference} and \ref{sec:operation} have been implemented within a real-time software architecture written in the Python programming language. The application layer is designed to be technology agnostic and could potentially be used with different sensing, communication, and processing technology. The software can be executed either directly on the Peregrine devices, described in Section~\ref{sub:hardware}, or as individual processes running on a central computer. The Peregrine software passes commands to and receives feedback from the Peregrine firmware described in Section~\ref{sub:firmware}. This flexibility allows the Peregrine system to operate as an \ac{nln} testbed in which algorithms can be tested, compared, and matured to the point where they can be run in real-time on small low-power devices.

\subsection{Firmware} 
\label{sub:firmware}

The Peregrine firmware provides an interface layer between the technology agnostic software application layer and the \ac{uwb} radio. The implementation of the firmware allows \ac{nln} to be in implemented in a robust and distributed manner. Peregrine's firmware consists of five major functions: pairwise ranging, neighbor discovery, channel sensing, channel sounding, and error handling. The firmware codes are written in C and are always executed locally on the Peregrine devices. It is designed to be robust to errors common in ad hoc networks, such as reception of unexpected messages, missing response messages, and packet collisions. The firmware written for Peregrine is unique in that it is written to support cooperative, multi-agent \ac{nln}.

The primary function of the Peregrine firmware is pairwise ranging. Ranging is performed by exchanging a series of locally timestamped messages between a pair of nodes, referred to as the initiator and responder. A unique feature of the Peregrine firmware is that agents can be both initiators and responders, allowing cooperative ranging. An agent is only an initiator when it is commanded to perform a range measurement; otherwise it is a responder, waiting for an initiation message from other agents. Anchors always act as responders. Once a ranging exchange is initiated between nodes $j$ and $k$ by agent $j$, the two nodes will only accept messages from each other until the exchange has ended. The exchange ends when a predetermined number of messages have been exchanged or when an error occurs. For each message, transmit and receive time stamps are exchanged and used to calculate a range measurement $z^{(n)}_{jk}$.\footnote{A single range measurement is calculated according to \cite{DW1000Manual:M} using the six timestamps generated from the first three, of four, messages. Using the first three messages guarantees that both initiator and responder have identical range estimates following the exchange. Using at least three messages makes the range measurement robust to both time and frequency differences on the two devices.} Range measurements are used for node inference in the software layer.

Neighbor discovery is performed automatically by the Peregrine firmware without any control input from the software layer. This is ensured with the addition of an auxiliary short broadcast message, called a ``chirp.'' Chirps are transmitted randomly by each node at a low average rate to ensure all nodes have knowledge of their neighbors, even if they aren't actively ranging. Because of the low transmission rate and short duration, the addition of the chirp message does not noticeably impact overall network channel usage. An agent updates its neighbor list based on the messages it receives from other nodes. Specifically, agent $j$ adds node $k$ to its neighbor list at time $n$ if it receives a message (including a chirp) from $k$ , i.e., $\Set{N}_j^{(n+1)} = \Set{N}_j^{(n)} \cup \{k\}$; agent $j$ also removes a node from its list of neighbors if a message from that node is not received for a certain period of time. The chirp from node $k$ also contains its state information. The neighbor set $\Set{N}_j^{(n)}$ and state information $\{\V{\mu}_{k}^{(n)}, \M{C}_{k}^{(n)}\}$ is provided to the software layer where it can be used by node inference and network operation algorithms.

Channel sensing is an additional function of the Peregrine firmware. When commanded to sense the channel for a specific time interval, a Peregrine node listens until a message is received or the time interval expires. If a message is received then the channel is immediately indicated as occupied, otherwise the channel is indicated as free at the end of the time interval. The channel sensing feature is needed to inform random access protocols, such as node activation algorithm, at the software layer.

Channel sounding is also enabled by the Peregrine firmware. If agent $j$ receives a message from node $k$, the channel impulse response can be measured and used to evaluate the quality of the channel between the two nodes. In particular, from the channel impulse response, the \ac{erc} $\erc_{jk}^{(n)}$ is estimated using a calibrated relationship to the waveform confidence metric described in \cite{APS006-3:M}. Channel sounding is performed during neighbor discovery and ranging. \acp{erc} $\erc_{jk}^{(n)}$, $k \in \Set{N}^{(n)}_j$ are used in the software layer by node inference, node activation, and node prioritization. 

\subsection{Hardware} 
\label{sub:hardware}

\begin{figure}[t]
	\centering
	\includegraphics[scale=.77]{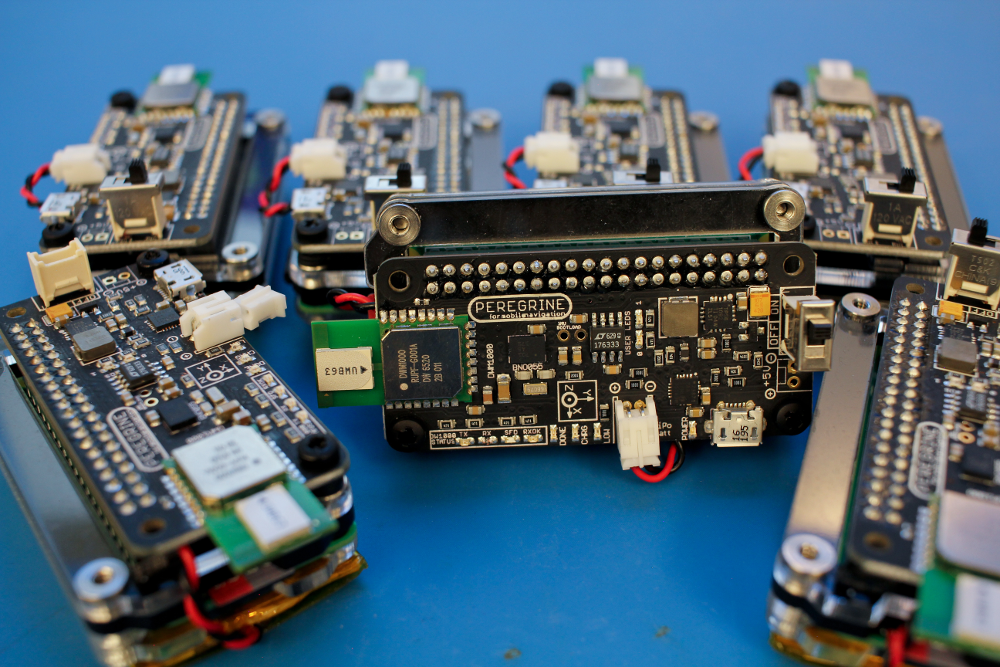}
	\caption[Peregrine node]{Devices used as nodes in Peregrine. A device is roughly the size of a business card. It includes a custom PCB with a \ac{uwb} radio module, a low-cost microprocessor, and a battery.}
	\label{fig:radios}
	\vspace{-1mm}
\end{figure}

The individual hardware nodes that compose the Peregrine system are designed to be fully self-contained sensing and processing devices. Each device, shown in Fig.~\ref{fig:radios}, comprises a microprocessor, a \ac{uwb} radio module, an WiFi radio, and a lithium-ion battery. The microprocessor can be simultaneously connected to the 802.11 wireless network and the \ac{uwb} localization network. 
Ranging and communication among Peregrine nodes is performed using the Decawave DWM1000, a commercially \ac{uwb} radio module. The Decawave \ac{ic} implements the IEEE 802.15.4 \ac{uwb} physical layer designed for \ac{lr-wpans}. The Peregrine system uses a module that includes the DW1000 \ac{ic} with a miniaturized \ac{uwb} antenna. With the ranging protocol described above, the ranging accuracy of this module is roughly 10cm, but it can be degraded in propagation environments which are strongly affected by multipath propagation. The radio module is integrated onto a \ac{pcb} designed specifically for the Peregrine node. The \ac{pcb} also contains power supply circuitry allowing it to use wall power or a standard lithium-ion battery cell.

The Peregrine \ac{pcb} is hosted on a Raspberry Pi Zero microprocessor board. The processing capability of this processor is roughly 1\% of that found in a modern Intel i7 processor. However, the Raspberry Pi Zero is capable of running the Peregrine firmware, executing \ac{nln} algorithms, and transmitting data on a 802.11 network for central processing or display.
\section{Performance Evaluation}
\label{sec:results}

A series of experiments were conducted to evaluate the performance of Peregrine. For all experiments, we use the \ac{cvm} as in \cite[Section 6.3.2]{ShaKirLi:B02} with state $\RV{x}_{j}^{(n)} = \big[\RV{p}^{(n) \ist \text{T}}_{j} ~ \ist \RV{v}^{(n) \ist \text{T}}_{j} \big]^{\text{T}}$, where $\RV{v}_{j}^{(n)} \in \mathbb{R}^{3}$ is the current 3-D velocity.\footnote{The \ac{cvm} can be trivially replaced with another linear motion model as the application requires.} In particular, $\M{A}({t^{(n)}})$ and $\RV{w}_{j}^{(n)}$ in \eqref{eq:cvm} are set as follows
\begin{align*}
	\M{A}({t^{(n)}}) &= 
		\begin{bmatrix}
			\M I_3 & t^{(n)} \M I_3 \\
			\M 0_{3 \times 3}  & \M I_3
		\end{bmatrix} \\[3mm]	    
	\RV{w}_{j}^{(n)} &= \M{B}({t^{(n)}}) \ist \RV{u}_{j}^{(n)} = 
		\begin{bmatrix} 
			\frac{1}{2}\big[t^{(n) }\big]^2 \ist \M I_3\\
			  t^{(n)} \M I_3
		\end{bmatrix} \ist \RV{u}_{j}^{(n)},   
\end{align*}
where the process noise $\RV{u}_{j}^{(n)}$ is distributed as $\RV{u}_{j}^{(n)} \rmv\sim \Set{N}(\V{0}, \M{C}_u)$ with $\M{C}_u=\text{bdiag}\{\sigma^2_x ~ \sigma^2_y ~ \sigma^2_z\}$. 
\begin{figure}[t]
	\centering
	\vspace*{-4mm}
	\includegraphics[scale=.9]{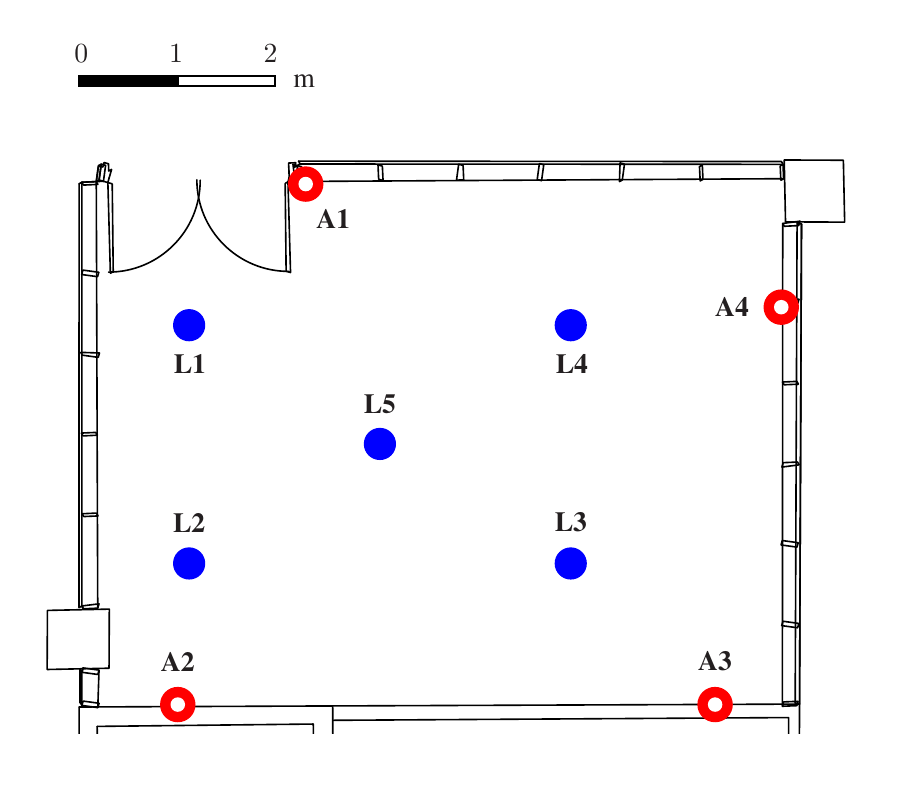}
	\vspace*{-5mm}
	\caption{Test environment of the single-floor experiments. Red circles indicate anchors A1 to A4 and blue dots indicate landmarks L1 to L5. (The floor plans used in this paper were provided by the Department of Facilities, Massachusetts Institute of Technology.)\vspace{0mm}}
	\label{fig:test}
\end{figure}
Important parameters are set as follows. The variance of the driving noise is set to $\sigma^2_{{x}} = \sigma^2_{{y}} = (0.06$m/s$^2)^2$, $\sigma^2_{{z}}  = (0.02$m/s$^2)^2$. In the measurement model, the variance of the measurement noise is set to $\big[\sigma^{(n)}_{jk}\big]^2 =  \big(\xi_{jk}^{(n)} \times m^{(n)}_{jk} \big)^{-1}$, where the \ac{erc} $\xi_{jk}^{(n)}$ varies from $16.0 \text m^{-2}$ to $100.0 \text m^{-2}$, depending on the waveform confidence metric.  Unless noted otherwise, we set the number of measurements performed with each neighbor to $\|\V{m}^{(n)}_j\| = 4$. This results in $t^{(n)}_j = T_m \times \| \V{m}^{(n)}_j \|$ for the \ac{htna} algorithm, where  $T_m$ is the channel access time related to performing a single measurement. The \ac{rmse} of the 3-D position estimate and the \ac{leo} were used as performance metrics. We performed the experiments in a single-floor scenario and a multi-floor scenario.

\subsection{Single-Floor Experiments}

The experiments took place in a rectangular room with metal walls, as shown in Fig.~\ref{fig:test}. Four static anchors, referred to as A1 through A4, are placed on the walls in the room. Five landmarks with known positions in the room, referred to as L1 through L5, were used as ground truth for evaluating localization accuracy. At all five landmarks, distance measurements suffer from the effects of multipath propagation. Agents are either static at a particular landmark, or move between landmarks in a predefined order.
The algorithms for node inference and network operation in Peregrine are compared with reference techniques to investigate their effects on localization performance and the measurement rate of the network. The reference node inference method is the \ac{ls} algorithm, the reference node activation methods include ALOHA \cite{Abr:B673} and \ac{csma} \cite{KriShi:B97}, and the reference for node prioritization is uniform allocation. For notational simplicity, we use acronyms to denote different combinations of node inference, node activation, and node prioritization methods, and these acronyms are listed in Table~\ref{tab:algsum}.

\subsubsection{Single Node Inference}

We investigate the performance of the \ac{spbp}-based algorithm for node inference in a network with a single agent. We compare the localization performance of the \ac{spbp} algorithm with that of the \ac{ls} algorithm as in \cite{Wan:J15} using Peregrine. The \ac{ls} algorithm applies the gradient descent procedure for determining the agent position, and the position estimate at the previous time $n-1$ is used as the start point for the gradient descent procedure at time $n$.
\begin{table}[t]
\centering
\vspace*{1mm}
\begin{tabular}{| c | C{0.9cm} | C{1.2cm} | C{1.4cm} | C{1.7cm} |}
\hline
& & \multicolumn{3}{c|}{Algorithm} \\
\cline{3-5}
Acronym & Section Used & Node Inference & Node Activation & Node Prioritization \\
\hline

\rowcolor{blue!10!white} 
LS-AL-UN & 6.1.1 & \ac{ls}  & ALOHA & Uniform \\
BP-AL-UN & 6.1.1-3 & \ac{spbp} & ALOHA  & Uniform \\
\rowcolor{blue!10!white} 
BP-CS-UN & 6.1.3 & \ac{spbp}  & CSMA & Uniform \\
BP-HT-UN & 6.1.3-4, 6.2 & \ac{spbp}  & \ac{htna} & Uniform \\
\rowcolor{blue!10!white} 
BP-HT-CP & 6.1.4, 6.2 & \ac{spbp} & \ac{htna} & \ac{cpnp} \\

\hline
\end{tabular}
\vspace{2.5mm}
\caption{Acronyms for different combinations of algorithms}
\label{tab:algsum}
\vspace{-6mm}
\end{table}

\begin{table}[ht]
\centering
\begin{tabular}{| C{1.4cm} | C{0.9cm} | C{0.9cm} | C{0.9cm} | C{0.9cm} | C{1.2cm} |}
\hline
& \multicolumn{5}{c| }{\ac{rmse} of the Position Estimate [cm]} \rule{0mm}{3.3mm}\\
\cline{2-6}
{}&{L1}&{L2}&{L3}&{L4}&{Average} \\
\hline

\rowcolor{blue!10!white} LS-AL-UN & 49   & 40   & 48  &  38   & $44 \ist\ist $ \\

BP-AL-UN & 43   & 33   &  37  & 23   & 34  \\

\hline
\end{tabular}
\vspace{2.5mm}
\caption{Single Node Inference Results: \acp{rmse} of the position estimates at each individual landmark and averaged over all landmarks for LS-AL-UN and BP-AL-UN. }
\label{tab:resultsInference}
\vspace{-6mm}
\end{table}
In this experiment, the agent moves between landmarks L1 and L4 in a defined order. Each landmark is visited ten times, and at each visit the estimate of the position is recorded for ten seconds. The unslotted ALOHA protocol \cite{Abr:B673} is adopted as the node activation method, and uniform allocation is adopted as the node prioritization method. In other words, LS-AL-UN and BP-AL-UN are considered. In uniform allocation, an agent randomly selects one of its neighbors according to a uniform distribution and makes measurements with it.

We evaluate the \acp{rmse} of the 3-D position estimates at landmarks L1, L2, L3, and L4, and compute the \ac{rmse} averaged over all the four landmarks. The results of LS-AL-UN and of BP-AL-UN are shown in Table~\ref{tab:resultsInference}. It can be seen that BP-AL-UN has a reduced \ac{rmse} at all landmarks compared to LS-AL-UN. In addition, the average \ac{rmse} of the position estimate is reduced by 23\% with BP-AL-UN compared to LS-AL-UN.

\subsubsection{Cooperative Node Inference}

We investigate the localization performance improvement from spatial cooperation in a network with two agents. Specifically, we compare the localization performance of Peregrine when two agents perform cooperative node inference with the case where the agents perform noncooperative node inference. In the latter scenario, the agents do not cooperate with each other and they only make distance measurements with the anchors.

Each experiment contains two phases. In the first phase, agent 1 is able to make measurements with all the four anchors, whereas agent 2 only makes measurements with three anchors. In the second phase, one of the three anchors both agents are connected to is switched off. As a result, agent 1 and agent 2 are able to make measurements with three and two anchors, respectively. Both phases last for ninety seconds. We did experiments under three geometric variations of the previously described experiment, referred to as configurations C1, C2, and C3.

We performed five independent experiments for each configuration. For both noncooperative and cooperative scenarios, BP-AL-UN is used, i.e., \ac{spbp}, unslotted ALOHA, and uniform allocation are adopted as the node inference, node activation, and node prioritization algorithms, respectively. 

\begin{figure}[t]
	\centering
	\vspace*{-.9mm}
	\includegraphics[width=0.95\columnwidth,draft=false,trim={0 0 0 0.35cm},clip]{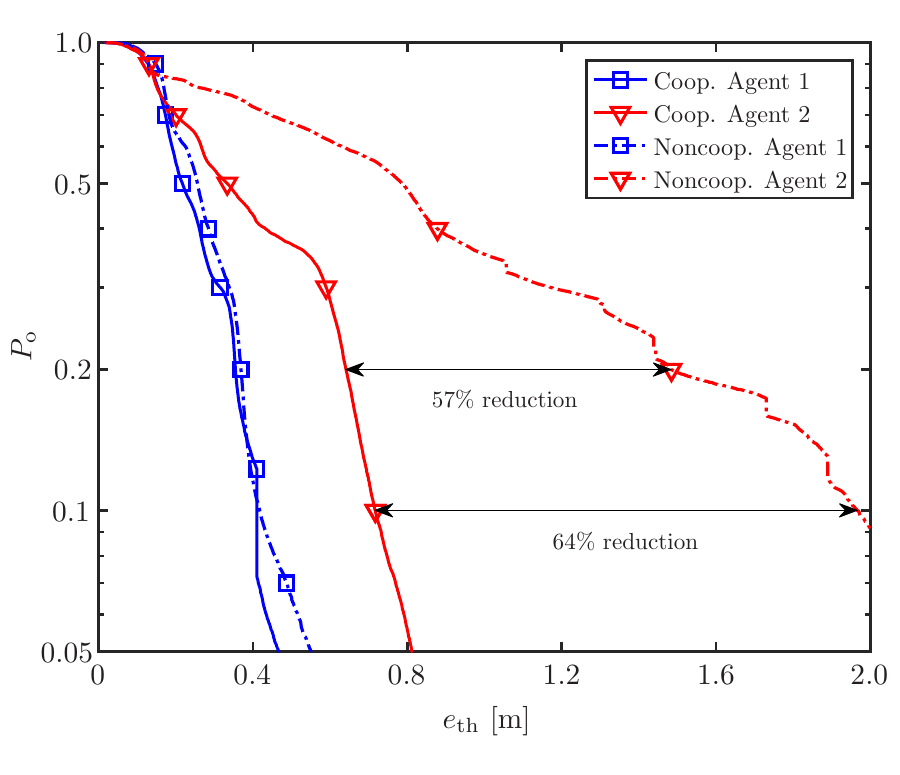}
	\vspace{-3.5mm}
	
	\caption{The \ac{leo} as a function of $e_{\text {th}}$ [m] in   noncooperative and cooperative scenarios.}
	\label{fig:coopError}
\vspace{2mm}
\end{figure}

We quantify the performance during each experiment in terms of the 3-D \ac{leo} $P_{\text o}$, defined in \cite{ConGueDarDecWin:J12} as the outage probability for a localization error threshold $e_{\text{th}}$.\footnote{The outage probability is a common performance metric for wireless communication systems (see, e.g., \cite{ConWinChiWint:L03}). The analogy with network localization is the probability that the quality of service falls below an acceptable level.} Fig.~\ref{fig:coopError} shows that for both agents, the \ac{leo} is improved in the cooperative scenario compared to the noncooperative scenario, and the improvement of agent 2 is much more significant than that of agent 1. 
Specifically, when \ac{leo} is 0.2, in the cooperative scenario, $e_{\text{th}}$ is reduced by $57\%$ from $1.48$m to $0.64$m for agent 2. Similarly, when \ac{leo} is 0.1, the $e_{\text{th}}$ is reduced by $64\%$ from $1.97$m in the noncooperative scenario to $0.71$m in the cooperative scenario. 
Agent 2, the agent with fewer anchor connections, benefits more from cooperation because it is left with only two anchor connections in the second stage of the experiment. With only two connections, agent 2 suffers from large error as shown in Fig.~\ref{fig:coopError}. When cooperation with agent 1 is enabled, a third connection is formed, the ambiguity is resolved, and the resulting error drops dramatically. Agent 1 is left with three anchor connections in stage 2, which is enough to unambiguously localize an agent given the high quality prior from stage 1. As a result, the accuracy is only slightly reduced with the loss of a fourth anchor.

Table~\ref{tab:resultsCooperation} shows the \ac{rmse} of the 3-D position estimate in C1, C2, and C3 as well as the resulting average \ac{rmse}. In particular, it can be seen that the \ac{rmse} of the position estimate average over all configurations is reduced by 53\% with cooperative node inference compared to noncooperative node inference.
\begin{table}[t]
\centering
\vspace*{1mm}
\begin{tabular}{| C{2.0 cm} | C{1.0cm} | C{1.0cm} | C{1.0cm} | C{1.2cm} |}
\hline
& \multicolumn{4}{c| }{\ac{rmse} of the Position Estimate [cm]} \\
\cline{2-5}
{}&{C1} & {C2} & {C3}& {Average} \\
\hline

\rowcolor{blue!10!white} 
Noncooperative & 67   & 85   & 103   & 85   \\
Cooperative & 31  & 33   & 56   & 40   \\

\hline
\end{tabular}
\vspace{2.5mm}
\caption{Node Inference Results: \acp{rmse} of the position estimate with and without cooperation.}
\label{tab:resultsCooperation}
\vspace{-6mm}
\end{table}

\subsubsection{Node Activation}
\label{sec:resultActivation}

We investigate the performance of various node activation methods in a network with three agents. In particular, we compare the localization performance of \ac{htna} with the performance of two reference protocols, namely unslotted ALOHA and \ac{csma} \cite{KriShi:B97}.

Table~\ref{tab:resultsActivation} shows the average \ac{rmse} of the position estimates and the average measurement rates for BP-AL-UN, BP-CS-UN, and BP-HT-UN. The measurement rate is averaged over the duration of each experiment. 
\begin{table}[t]
\centering
\begin{tabular}{ | C{1.5cm} | C{2.7cm} | C{2.5cm} |}
\hline
{}&{\ac{rmse} of the Position Estimate [cm] }&{Measurement Rate [Hz] } \\
\hline

\rowcolor{blue!10!white} 
BP-AL-UN & $41 $ & $41$ \\
BP-CS-UN & $33 $ & $74$ \\

\rowcolor{blue!10!white} 
BP-HT-UN & $34 $ & $40$ \\

\hline
\end{tabular}
\vspace{2.5mm}
\caption{Node Activation Results: \acp{rmse} of the position estimate and measurement rates of the network for different node activation strategies. }
\label{tab:resultsActivation}
\vspace{-6mm}
\end{table}
We make the following observations from Table~\ref{tab:resultsActivation}. The \ac{rmse} of the position estimates of BP-HT-UN is similar to that of BP-CS-UN, but the measurement rate reduced by $46\%$. The reason is that only agents with channel access indicator being one, as described in Section~\ref{sec:nodeActivation}, will attempt to access the channel in BP-HT-UN. As a result, only the subset of agents that will benefit the most will attempt to access the channel. This demonstrates that the channel is efficiently shared between the agents in a manner that benefits localization. Comparing BP-AL-UN and BP-CS-UN we see that given a similar measurement rate, location-aware node activation results in better position accuracy than pure random access because it is granting more frequent channel access to the nodes that will benefit most.

\subsubsection{Node Prioritization}

We evaluate the performance of the \ac{cpnp} algorithm in a network with a single agent. In particular, we compare the localization performance of the \ac{cpnp} algorithm with a uniform allocation.

In these experiments, a static agent is placed at landmark L5. In order to create an indoor environment with severe multipath effects, we placed a large metal plate near landmark L4. We performed five independent experiments for both methods, each with a duration of two minutes. The \ac{spbp} algorithm and \ac{htna} method are adopted for node inference and node activation, respectively, i.e., the BP-HT-UN and BP-HT-CP techniques are considered. For BP-HT-CP, we fix the total number of distance measurements an agent can make at each time slot to $M_j^{(n)} = 12$ (see Section~\ref{sec:nodePrioritization}).
\begin{table}[t]
\centering
\vspace*{1mm}
\begin{tabular}{| C{1.4cm} | C{0.9cm} | C{0.9cm} | C{0.9cm} | C{0.9cm} |}
\hline

 & \multicolumn{4}{c| }{\ac{rmse} of Distance Measurements [cm]} \\
\cline{2-5}
&{A1} & {A2} & {A3} & {A4} \\
\hline
\rowcolor{blue!10!white} 
All & $10.6$ & $19.7$ & $13.2$ & $50.1$ \\

\hline

& \multicolumn{4}{c| }{Fraction of Distance Measurements [\%]} \\
\cline{2-5}
\hline
\rowcolor{blue!10!white}
BP-HT-UN & $24.8$ & $25.1$ & $24.8$ & $25.2$ \\
BP-HT-CP & $48.8$ & $15.9$ & $25.6$ & $9.7$ \\
\hline
\end{tabular}
\vspace{2.5mm}
\caption{Node Prioritization Results: \ac{rmse} and fraction of the distance measurements that the agent performed with each anchor.}
\label{tab:freqPriori}
\vspace{-6mm}
\end{table}
Our results show that with BP-HT-UN, the \ac{rmse} of the 3-D position estimates is $43$cm, and such error is reduced to $38$cm with BP-HT-CP. We verified the accuracy of the distance measurements performed by the single agent with all anchors. As shown in Table~\ref{tab:freqPriori}, the \acp{rmse} of the distance measurements performed with anchors A2 and A4 are significantly larger than those performed with anchors A1 and A3. Table~\ref{tab:freqPriori} also shows the fraction of distance measurements the agent makes with each anchor. It can be seen that with uniform allocation, the agent makes an approximately equal number of measurements with each neighbor, whereas with the \ac{cpnp} algorithm, the agent makes more measurements with anchors A1 and A3 than with anchors A2 and A4. This is because Peregrine is able to detect that the received signals from anchors A2 and A4 are heavily impacted by multipath effects, and thus the \acp{erc} $\xi_{jk}^{(n)}$'s are set to small values for these anchors. As a result, the \ac{cpnp} algorithm gives lower priority to anchors A2 and A4, and the agent performs fewer distance measurements with anchors A2 and A4 compared to anchors A1 and A3.

\subsection{Multi-Floor Experiment}

\begin{figure*}[t]
\centering

        		\includegraphics[width=0.91\textwidth,draft=false]{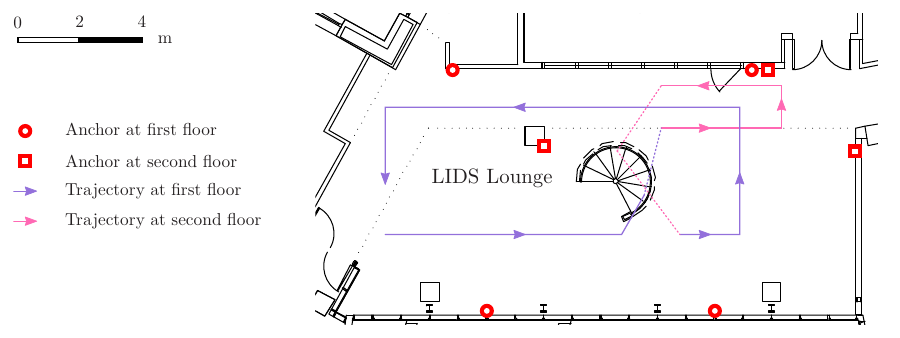}%

\caption[]{Multi-floor test environment: four anchors (red circles) and three anchors (red squares) are placed on the walls of the
first and second floor, respectively. An agent walks along the trajectory that spans both the first floor (purple line) and the second floor (magenta
line).}
\label{fig:lounge_config}
\end{figure*}

We evaluate the performance of the Peregrine system in single-agent, multi-floor scenarios. The goal is to characterize the performance of Peregrine in a larger, more complicated, and multipath-dense environment as well as to demonstrate performance gains enabled by the node prioritization algorithm. The experiments took place in the lounge of \ac{lids}, which is an open space consisting of two floors (see Fig.~\ref{fig:lounge_config}). We placed seven anchors in the environments: four are attached to the walls on the first floor, and three are attached to the walls of the second floor. An agent walks along a trajectory across both floors. On the trajectory we placed twenty-three landmarks with equal spacing: sixteen on the first floor and seven on the second floor. The agent walks along the trajectory and stops at each landmark for thirty seconds. During this time, 3-D position estimates are recorded and the 3-D localization error of Peregrine is calculated.

\begin{figure}[t]
	\centering
	\vspace*{-.9mm}
	\includegraphics[width=0.95\columnwidth,draft=false,trim={0 0 0 0.35cm},clip]{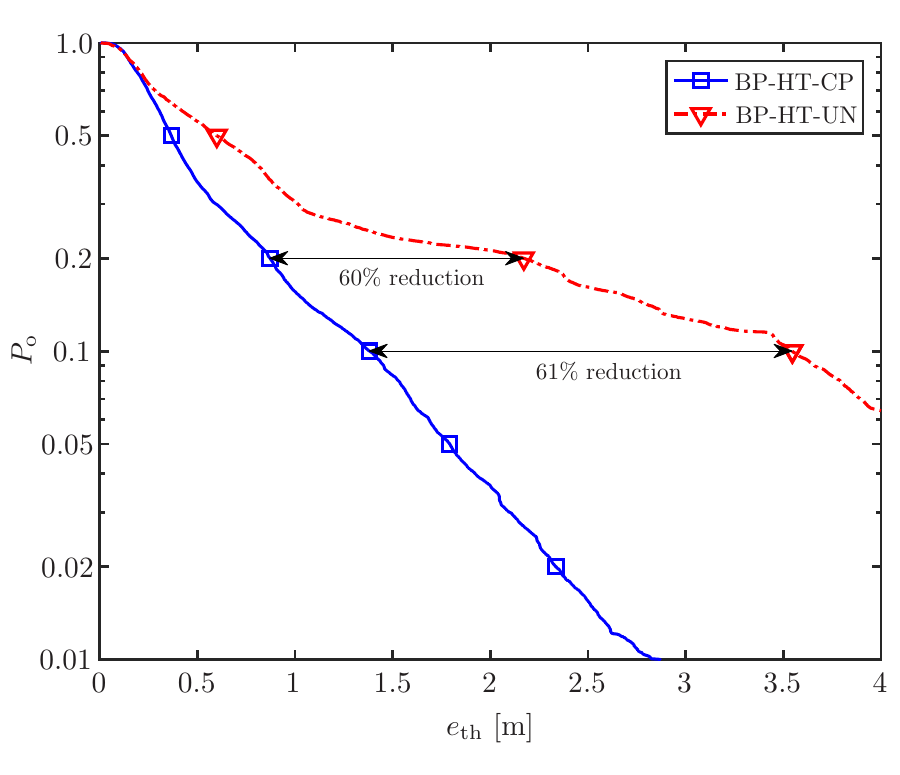}
	\vspace{-3.5mm}
	
	\caption{The \ac{leo} as a function of $e_{\text {th}}$ [m] for BP-HT-CP and BP-HT-UN in multi-floor test environments}
	\label{fig:leo-lounge}
\vspace{2mm}
\end{figure}

We performed three independent experiments and considered the BP-HT-UN and BP-HT-CP techniques. The 3-D \ac{leo} metrics are shown in Fig.~\ref{fig:leo-lounge}. It can be seen that the localization performance is significantly improved with BP-HT-CP. Specifically, when the \ac{leo} is 0.2, the $e_{\text{th}}$ is reduced by $60\%$ from $2.17$m with BP-HT-UN to $0.87$m with BP-HT-CP. Similarly, when \ac{leo} is 0.1, the $e_{\text{th}}$ is reduced by $61\%$ from $3.55$m with BP-HT-UN to $1.38$m with BP-HT-CP. The localization performance degradation with BP-HT-UN mainly results from the biases in the range measurements from NLOS agent-anchor links. In particular, a significant proportion of the anchors, especially the ones not at the same floor as the agent, are in severe \ac{nlos} with respect to the agent. 
As a result, the measurements performed by the agent with anchors in \ac{nlos} contain large biases and can potentially degrade the localization performance. BP-HT-UN performs an approximately equal number of measurements with anchors in both \ac{los} and \ac{nlos}, evaluates the channel quality with each anchor, and adjusts the noise variance in the measurement model as discussed in the beginning of this section. However, compared to BP-HT-BP, the localization performance is still degraded, since with BP-HT-UN a substantial amount of resources is used to perform measurements with \ac{nlos} anchors. In contrast, with BP-HT-CP, the agent focuses on performing measurements with anchors in \ac{los}. These measurements do not suffer from significant biases and thus the localization performance is improved. This set of experiments confirms the reliability of Peregrine in harsh radio propagation environments enabled by efficient node prioritization.

\acresetall

\section{Conclusion}\label{sec:conclusion}
This paper introduced Peregrine, the first system for 3-D cooperative \ac{nln}. The devices used as nodes in Peregrine include a ultra-wideband radio module as well as a low-cost microprocessor. They are self-contained, low-cost, and compact. Peregrine demonstrates a complete architecture for \ac{nln}, featuring state-of-the-art algorithms for scalable node inference and efficient network operation. We quantified the benefits of spatial cooperation and of the node inference, node activation, as well as node prioritization algorithms with experimental results. These results demonstrate that Peregrine achieves reliable localization performance with efficient resource utilization even in harsh radio propagation environments.

\ifCLASSOPTIONcaptionsoff
  \newpage
\fi



%

\bibliographystyle{IEEEtran}
\bibliography{Bib/IEEEabrv,Bib/StringDefinitions,Bib/BiblioCV,Bib/Wgroup,Bib/Temp,Bib/WINS-Books}

%
%

%

%






\end{document}